\documentclass{emulateapj}

\usepackage{float}
\usepackage{amsmath}
\usepackage{epsfig,floatflt}
\usepackage{subfigure}

\begin{document}

\title{Cleaned Three-Year WMAP CMB Map: Magnitude of the Quadrupole and 
       Alignment of Large Scale Modes}

\author{Chan-Gyung Park, Changbom Park}\affil{Korea Institute for Advanced Study, Seoul 130-722, Korea}
%\email{parkc@kias.re.kr;cbp@kias.re.kr}
\and \author{J. Richard Gott III}\affil{Department of Astrophysical Sciences,
Peyton Hall, Princeton University, Princeton, NJ 08544-1001, USA}

%\author{???}
%\affil{???}
%\email{???}
%\date{Received March 1, 2004 / Accepted May 20, 2004}

\begin{abstract}
We have produced a cleaned map of the \emph{Wilkinson Microwave Anisotropy 
Probe} (WMAP) three-year data using an improved foreground subtraction 
technique. We perform an internal linear combination (ILC) to subtract
the Galactic foreground emission from the temperature fluctuations observed 
by the WMAP.
We divide the whole sky into hundreds of pixel groups with similar foreground 
spectral indices over a range of WMAP frequencies, apply the ILC for each 
group, and obtain a CMB map with foreground emission effectively reduced.
With the resulting foreground-reduced ILC map (Figure \ref{fig:SILC400}$b$, 
available on-line), we have investigated the known anomalies in CMB maps 
at large scales, namely the low quadrupole ($l=2$) power, and the strong 
alignment and planarity of the quadrupole and the octopole ($l=3$).
Our estimates are consistent with the previous measurements.
The quadrupole and the octopole powers measured from our ILC map are
$\delta T_2^2 = 276_{-126}^{+94}~\mu\textrm{K}^2$ and
$\delta T_3^2 = 952_{-83}^{+64}~\mu\textrm{K}^2$, respectively.
The 68\% confidence limits are estimated from the ILC simulations and
include the cosmic variance.
The measured quadrupole power is lower than the value expected 
in the concordance $\Lambda$CDM model ($1250~\mu\textrm{K}^2$),
in which the probability of finding a quadrupole power lower than the measured 
value is $5.7$\%.
We have confirmed that the quadrupole and the octopole are strongly aligned
with angle $\theta_{23} = 11\fdg8_{-8\fdg0}^{+6\fdg4}$, and are planar with
high planarity parameters $t=0.98_{-0.02}^{+0.02}$ for $l=2$ and 
$t=0.91_{-0.03}^{+0.02}$ for $l=3$.
The observed angular separation $\theta_{23}$ is marginally statistically
significant because the probability of finding the angular separation 
as low as the observed value is 4.3\%. 
However, the observed planarity is not statistically significant. 
The probability of observing such a planarity as high as the measured $t$ 
values is over 18\%.
The ILC simulations show that the residual foreground emission in the ILC map
does not affect the estimated values significantly.
The large scale modes ($l=2$--$8$) of SILC400 shows anti-correlation 
with the Galactic foreground emission on the southern hemisphere. 
It is not clear whether such anti-correlation occurs due to the residual 
Galactic emission or by chance.
\end{abstract}

\keywords{cosmic microwave background --- cosmology: observations ---
methods: numerical}

\maketitle

\section{Introduction}
\label{sec_introduction}

The \emph{Wilkinson Microwave Anisotropy Probe} (WMAP; \citealt{ben03:basic}) 
has measured the cosmic microwave background (CMB) temperature anisotropy 
and polarization with high resolution and sensitivity, and opened a new 
window to precision cosmology. The WMAP one-year data imply that the observed 
CMB fluctuations are consistent with predictions of the concordance 
$\Lambda\textrm{CDM}$ model with scale-invariant adiabatic fluctuations 
generated during the inflationary epoch \citep{hin03,kog03,spe03,pag03,pei03}.
The recent release of the WMAP three-year data has confirmed the primary 
results from the one-year data, giving more accurate determination of 
cosmological parameters \citep{hin06,spe06,pag06}.

However, some peculiar aspects of the CMB maps have been noticed, leading
to many controversies. First, there have been many reports of detection
of non-Gaussian signatures (\citealt{chi03,par04,eri04:asym,eri04:genus,
col04,cop04,vie04,cru05,cru06,toj06}), as opposed to the WMAP team's result 
\citep{kom03,spe06}. 
In particular, the origin of the asymmetry in statistical properties 
of the CMB between the Galactic northern and southern hemispheres 
(\citealt{par04,eri04:asym}) has not been fully explained. 
\citet{fre06} have studied the effects of the map-making algorithm 
on the observed asymmetry in CMB temperatures between the Galactic northern 
and southern hemispheres in the WMAP data. 
Another interesting issue is the low CMB quadrupole power.
Cut-sky analysis of WMAP three-year data gives a quadrupole power of
$211~\mu\textrm{K}^2$, which is quite a bit lower than the expected value
in the best-fit $\Lambda\textrm{CDM}$ model ($1250~\mu\textrm{K}^2$;
\citealt{hin06,spe06}).
Through subsequent analyses of the WMAP data, such a low quadrupole power
has been confirmed and the effect of the residual foreground emission on
the statistical properties of CMB has been extensively studied
(e.g., \citealt{teg03,efs04,eri04:ilc,bie04,bie05,slo04a,slo04b,nas06,deo06}).
The alignment and planarity of $l=2$ and $3$ modes on the sky are also
interesting features that have been seen in the WMAP data 
\citep{deo04,sch04,lan05,cop06,abr06}.

Although each of the anomalies in the CMB anisotropy may have its own 
cosmological origin, there is a possibility that the residual Galactic 
foreground emission has affected the nature of the observed temperature 
fluctuations. Therefore, foreground subtraction is very important as a 
starting point of all CMB-related analyses.
The most popular method of foreground removal is to model the Galactic
emission as the weighted sum of foreground templates such as synchrotron,
free-free, and dust emission maps (e.g., \citealt{ben03:galaxy}).
Another method of foreground removal is the internal linear combination
(ILC) method using multi-frequency data \citep{bra94,teg96,ben03:galaxy}.

The WMAP team has created the ILC map (hereafter WILC1YR) by computing 
a weighted combination of the WMAP maps that have been band averaged 
within each of the five WMAP frequency bands, all smoothed to $1\degr$ 
resolution \citep{ben03:galaxy}.
The weights are determined by minimizing the variance of temperatures 
in the combined map, with a constraint that the sum of the weights is one.
They have defined twelve disjoint regions on the sky, within which weights
are determined independently. Except for the biggest Kp2 region,
the other 11 regions are defined by subdividing the inner Galactic plane
with lines of constant Galactic longitudes.
However, the combination weights found by the WMAP team do not give the 
minimum variance, and their method of `non-linear searching' for the set 
of weights has not been described.

\citet{eri04:ilc} have improved the WMAP team's ILC method by applying 
a Lagrange multiplier to produce a variance-minimized ILC map
(hereafter LILC). They use the same 12 disjoint regions and the smoothing
scheme to remove discontinuous boundaries as used in \citet{ben03:galaxy}.
\citeauthor{eri04:ilc} has emphasized that the effects of noise on the 
performance of the ILC method is very important. If noise is high, the ILC 
method finds the best combination weights that minimize the instrumental 
noise rather than foregrounds.

\citet{teg03} have applied a variant of the ILC method to make a 
foreground-cleaned CMB map (hereafter TCM) with high resolution.
By noticing that the Galactic foregrounds appear dominantly on larger 
angular scales while the instrument noise dominates only at smaller scales,
they have applied linear combinations in harmonic space to remove foreground 
emission at different angular scales separately. Based on levels of the 
Galactic foreground intensity, \citet{teg03} have defined nine disjoint 
regions, where foreground removal has been done independently. 
Recently, \citet{deo06} have made a new foreground-cleaned CMB map 
(hereafter TCM3YR) by applying the same technique to the WMAP three-year data.

In the three-year data analysis, the WMAP team has made a new ILC map
(hereafter WILC3YR) by combining the WMAP data at five different bands
\citep{hin06}. From the one-year version of region definition map,
they have eliminated the Taurus A region that is too small to give 
a reliable foreground removal while they have added a new region to minimize 
the dust residuals in the Galactic plane (denoted as Region 1 in Fig. 
\ref{fig:WILC12}$a$).
They have used the Lagrange multiplier method as used in \citet{eri04:ilc}
to determine the ILC coefficients. To quantify the bias due to the residual
foreground emissions, they have performed one hundred Monte Carlo simulations
using the one-year Galaxy model based on the Maximum Entropy Method (MEM).
The three-year ILC map, WILC3YR, has been produced by subtracting this bias 
prediction.

All ILC methods require several disjoint regions that contain map pixels
with similar properties. Otherwise, the efficiency of the ILC method becomes 
very low, leaving significant residual foregrounds. The Galactic foreground 
emission appears as the sum of the individual emissions with different physical
natures, and its spectral behavior varies over a wide range of frequencies 
and over the whole sky. The disjoint regions defined by the WMAP team 
and \citet{teg03} do not fully reflect the Galactic foreground properties.

In this paper, we derive a new foreground-reduced CMB map by applying
the minimum-variance ILC method including information about the foreground 
spectral properties. The outline of this paper is as follows.
In $\S2$, we describe our simple minimum-variance ILC method, and present
a new definition of sky regions for ILC. In $\S3$, we derive a new 
foreground-cleaned CMB map from the WMAP three-year data.
Measurements of statistics that are related to the quadrupole and the octopole 
are given in $\S4$. We discuss our results in $\S5$.

Throughout this paper, the power spectrum is calculated by
\begin{equation}
   \delta T_l^2 = l(l+1) C_l / 2\pi,
\label{eq:power}
\end{equation}
with
\begin{equation}
   C_l = (2 l + 1)^{-1} \sum_{l=-m}^{m} |a_{lm}|^2,
\label{eq:cl}
\end{equation}
where $a_{lm}$ are the spherical harmonic coefficients in the Galactic
coordinate system.

\begin{figure*}
%\plotone{s1.eps}
\mbox{\epsfig{file=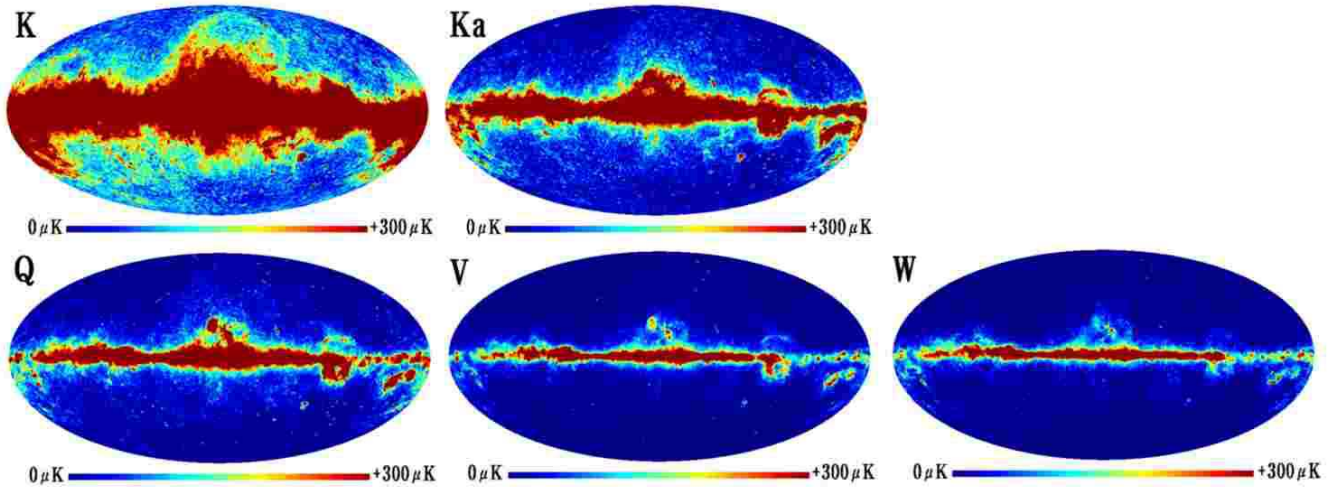,width=180mm,clip=}}
\caption{The three-year version of Galactic foreground emission maps 
         at five WMAP frequencies (K, Ka, Q, V, and W bands) made 
         by co-adding synchrotron, free-free, and dust emission components 
         derived from the MEM. The resolution of each map is $1\degr$ FWHM. 
         The Mollweide projection in Galactic coordinates is used to display 
         all maps. In each figure, the Galactic center is located at the center
         and the Galactic plane lies at the horizontal line that passes
         through the Galactic center. Galactic longitude increases from center 
         to left ($l=0\degr$--$180\degr$) and from right to center 
         ($l=180\degr$--$360\degr$) while Galactic latitude increases from 
         bottom to top ($b=-90\degr$--$+90\degr$).}
\label{fig:MEMG}
\end{figure*}

\section{Internal Linear Combination Method}
\label{sec_ilc}

\subsection{ILC with a Minimum Variance Constraint}
\label{sec_ilcmethod}

We find an optimal weight set for linear combination of observed CMB maps
to remove the Galactic foregrounds with the constraint that the combined
map have the minimum variance.
Let us denote the observed temperature at $p$-th pixel and at $i$-th frequency
band by $T_i(p)$. The observed temperature is a sum of the CMB temperature,
instrument noise, and the Galactic emission.
We perform ILC on a region $\mathcal{R}$ by minimizing the variance of the 
weighted sum of observed temperatures at pixels within the region.
For multi-frequency CMB data with $k$ bands, the temperature of $p$-th pixel
of the linearly combined map can be written as
\begin{equation}
   T(p) =\sum_{i=1}^{k} w_i T_i(p) = \sum_{i=1}^{k-1} w_i D_{ik}(p) + T_k(p),
\label{eq:ilc_temp}
\end{equation}
where $D_{ik}(p) \equiv T_i(p)-T_k(p)$ and $\sum_{i=1}^{k}w_i = 1$.
The variance of $T(p)$ over pixels within the region $\mathcal{R}$ is given by
\begin{equation}
\begin{split}
   \sigma^2 &(T) = {1 \over N_{\mathcal{R}}}
                  \sum_{p \in \mathcal{R}} \left[ T(p)-\bar{T} \right]^2  \\
       &={1 \over N_{\mathcal{R}}} \sum_{p \in \mathcal{R}}
          \left[ \sum_{i=1}^{k-1} w_i \left\{ D_{ik}(p)-\bar{D}_{ik} \right\}
          + \left\{T_k(p) - \bar{T}_k\right\} \right]^2,
\end{split}
\label{eq:var_temp}
\end{equation}
where $\bar{T}$ represents the average of $T(p)$ over $N_{\mathcal{R}}$ 
pixels within the specified region $\mathcal{R}$. ILC coefficients that 
give the minimum variance is obtained by solving linear equations 
$\partial \sigma^2(T) / \partial w_j = 0$, where $j=1,\cdots,k-1$.
The solution $\mathbf{w}=(w_1,\cdots,w_{k-1})^T$ is simply obtained from
\begin{equation}
   \mathbf{w} = -\mathbf{S}^{-1} \mathbf{t}.
\label{eq:w}
\end{equation}
Here components of the $(k-1)\times(k-1)$ symmetric matrix $\mathbf{S}$ and
the $(k-1)\times 1$ vector $\mathbf{t}$ are given by
\begin{equation}
\begin{split}
   &S_{ij}={1 \over N_{\mathcal{R}}} \sum_{p \in \mathcal{R}}
                  \left\{D_{ik}(p)-\bar{D}_{ik}\right\}
                  \left\{ D_{jk}(p)-\bar{D}_{jk} \right\}, \\
   &t_j = {1 \over N_{\mathcal{R}}} \sum_{p \in \mathcal{R}}
                  \left\{T_k(p)-\bar{T}_k \right\}
                  \left\{D_{jk}(p)-\bar{D}_{jk} \right\},
\end{split}
\label{eq:st}
\end{equation}
where $i,j=1,\cdots,k-1$. The weight for the last frequency band map is
$w_k = 1-\sum_{i=1}^{k-1} w_i$.
With the set of ILC weights, a foreground-reduced CMB map can be obtained 
from equation (\ref{eq:ilc_temp}) (see $\S$\ref{sec_appl}).

\subsection{Defining Pixel-Groups with Common Foreground Spectral Properties}
\label{sec_fore}

The efficiency of foreground reduction by the linear combination method 
becomes highest when the region contains pixels with the same foreground 
spectral variation over frequencies. 
On the other hand, if a region contains a large number of pixels whose 
Galactic foreground spectral property varies dramatically over the sky, 
as is the region $\mathcal{R}=0$ of WILC3YR (see Fig. \ref{fig:WILC12}$a$), 
the ILC just tries to find the best-fit combination weights that is optimal 
only to the average foreground spectral behavior. The ILC then results 
in positive or negative residual bias in the map ($\S\ref{subsec_applwmap}$).

To assess the Galactic foreground properties, we use the MEM Galactic 
foreground maps derived by the WMAP team, which are shown in Figure 
\ref{fig:MEMG}. The WMAP team has tried to distinguish different emission 
sources from one another by applying the MEM to the WMAP data, where the prior
spatial distribution and spectral behavior of foreground components have been 
assumed by using the Galactic template maps. From this information the team 
has produced synchrotron, free-free, and dust emission maps with $1\degr$ 
resolution at each WMAP frequency. For each band, we co-add three emission 
maps to make a Galaxy foreground map, upgrade its resolution to
$\textrm{N}\textrm{side}=512$, and further smooth the map with a Gaussian 
filter with variable widths ($0\degr$ -- $2\degr$) to reduce the MEM 
reconstruction noise. We smooth each Galaxy map over $\textrm{FWHM}=0\fdg5$, 
and then subtract it from the unsmoothed Galaxy map to make a difference map, 
from which the standard deviation is calculated at the region with 
$|b|>60\degr$. 
The signal-to-noise ratio at each pixel for each band is defined as the 
foreground intensity divided by the standard deviation. The minimum value 
among the five signal-to-noise ratios at each pixel is considered as the 
representative signal-to-noise ratio (S/N). The width of the smoothing filter 
has been set to zero for
$\textrm{S/N} \ge 10$, and $\textrm{FWHM} = 0.2(10 - \textrm{S/N})$ 
for $\textrm{S/N} < 10$.

Table \ref{tab:Galfore} lists the spherical harmonic coefficients ($a_{lm}$)
of the K-band Galactic foreground map up to $l=6$. The $a_{lm}$'s of 
foreground emissions at higher frequency bands have similar patterns with 
smaller amplitudes. The Galactic foreground emission is strong at even 
$l$-modes, among which $l=2$ and $m=0$ mode is the strongest one.
The real components of even (odd) $m$-modes for even (odd) $l$ are stronger 
than the imaginary components. Modes of strong Galactic signal are expected 
to affect the corresponding modes of CMB signal significantly, resulting 
in biases in the reconstruction maps.

\begin{deluxetable}{ccrrccrr}
\tablewidth{0pt}
%\tablewidth{220pt}
\tabletypesize{\small}
\tablecaption{Low $l$-mode spherical harmonic coefficients of the K-band 
              Galactic foreground emission \label{tab:Galfore}}
\tablecolumns{8}
\tablehead{$l$ & $m$ & Re[$a_{lm}$] & Im[$a_{lm}$] & $l$ & $m$ & Re[$a_{lm}$] & Im[$a_{lm}$]}
\startdata
2 & 0 & $-2153.7$  & $0.0$    &  5 & 0 & $1.6$    & $0.0$\\
  & 1 & $-140.0$   & $36.6$   &    & 1 & $-840.7$ & $40.5$\\
  & 2 & $412.5$    & $172.6$  &    & 2 & $2.2$    & $-56.4$ \\
3 & 0 & $14.0$     & $0.00$   &    & 3 & $158.2$  & $100.0$\\
  & 1 & $875.4$    & $-31.1$  &    & 4 & $3.5$    & $9.3$\\
  & 2 & $3.1$      & $41.4$   &    & 5 & $-107.8$ & $32.3$\\
  & 3 & $-218.3$   & $-115.8$ &  6 & 0 & $-1672.5$& $0.0$ \\
4 & 0 & $1871.2$   & $0.0$    &    & 1 & $-98.1$  & $100.1$\\
  & 1 & $121.2$    & $-97.2$  &    & 2 & $187.2$  & $120.3$\\
  & 2 & $-244.5$   & $-121.5$ &    & 3 & $66.1$   & $13.3$\\
  & 3 & $-104.9$   & $2.2$    &    & 4 & $-208.8$ & $13.5$     \\
  & 4 & $361.4$    & $-46.1$  &    & 5 & $-14.5$  & $33.7$  \\
  &   &          &            &    & 6 & $-208.6$ & $-121.5$ \\[-2mm]
\enddata
\tablecomments{The coefficients are calculated in Galactic coordinates
               in units of $\mu\textrm{K}$.}
\end{deluxetable}

\begin{figure}
\mbox{\epsfig{file=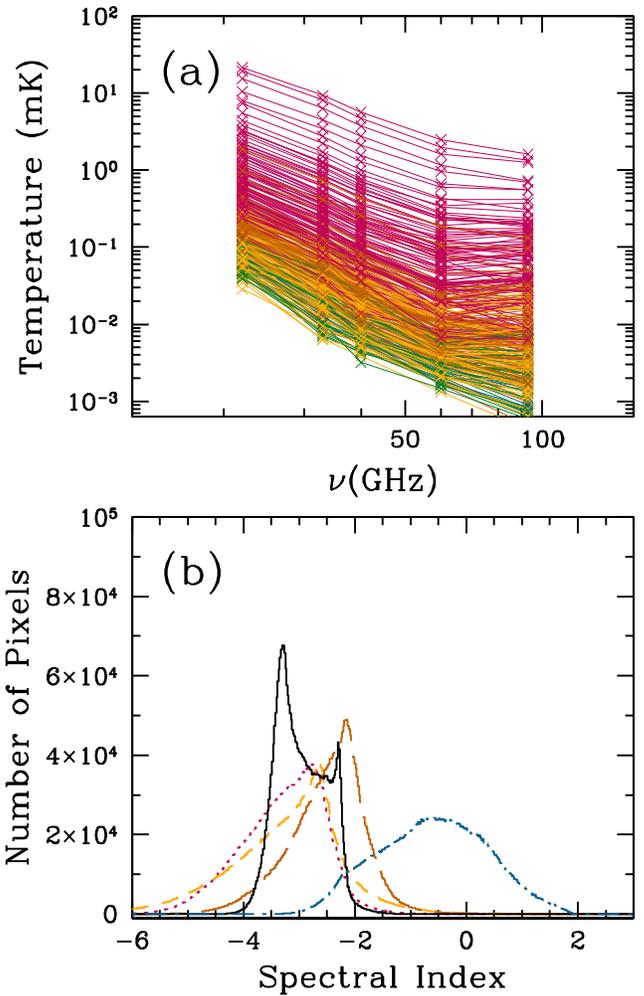,width=88mm,clip=}}
\caption{($a$) Intensity variation of the Galaxy foregrounds over WMAP 
         frequencies ($22.8$ [K], $33.0$ [Ka], $40.7$ [Q], $60.8$ [V], 
         and $93.5$ GHz [W]). According to levels of Galactic latitudes
         (low to high), plots are shown with different colors (red to blue). 
         Only about 200 pixel values are shown to avoid overlaps. 
         ($b$) Histograms of foreground spectral indices measured at K--Ka 
         (dotted), Ka-Q (dashed), Q--V (long-dashed), and V--W (dot-dashed 
         curve) frequency intervals. The solid curve denotes a histogram 
         of spectral indices measured from four foreground intensities at K 
         to V bands. The MEM Galactic foreground maps smoothed with a Gaussian 
         filter of variable widths have been used for spectral index 
         measurement.}
\label{fig:specdistr}
\end{figure}

Figure \ref{fig:specdistr}$a$ shows intensity variations of the Galactic
foreground emission across WMAP bands at about 200 randomly selected pixels.
Figure \ref{fig:specdistr}$b$ shows the histograms of foreground spectral
indices ($\alpha$) within K--Ka, Ka--Q, Q--V, and V--W frequency intervals
measured at all pixels in the sky. Each spectral index is measured by modeling 
the foreground intensity by $I_\nu \propto \nu^\alpha$.
The spectral indices measured at frequency interval V--W (dot-dashed curve)
have a wide range of values from negative to positive, centered at 
$\alpha\approx -0.6$ unlike those at K--Ka, Ka--Q, and Q--V that are located 
at $\alpha \lesssim -1$ (dotted, dashed, and long-dashed curves). Therefore, 
it is reasonable to treat spectral indices at low (K--V) and high (V--W) 
frequency intervals separately.
For simplicity, we use an average spectral index measured for four foreground
intensities from K to V bands as the representative spectral index at low 
frequency bands (Fig. \ref{fig:specdistr}$b$; solid curve). Hereafter we call 
the spectral indices measured at the low and high frequency intervals $\alpha$
and $\beta$, respectively.

\begin{figure}
\mbox{\epsfig{file=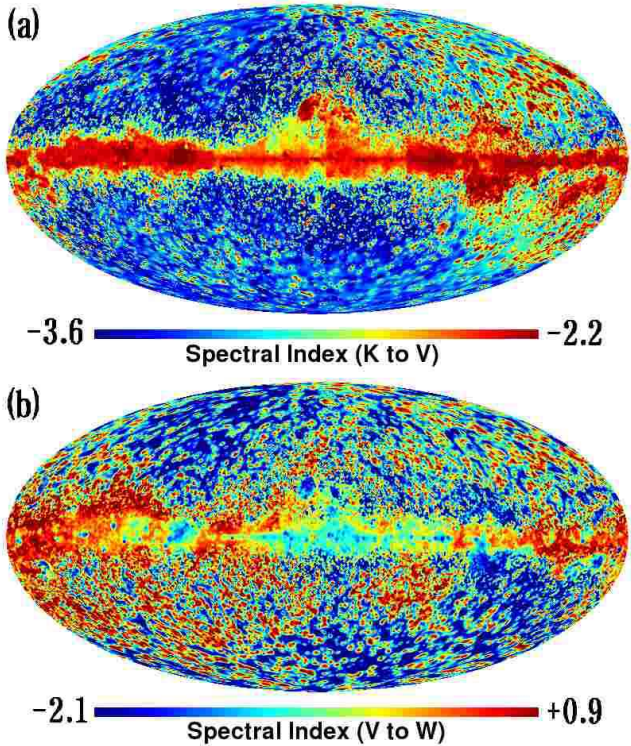,width=88mm,clip=}}
%\resizebox{\textwidth}{!}{\includegraphics{f3.eps}}
\caption{All-sky maps showing distributions of foreground spectral indices 
         measured at ($a$) low (K--V) and ($b$) high (V--W) frequency intervals.
         The MEM Galactic foreground maps smoothed with a Gaussian filter
         of variable widths have been used to measure the spectral indices.}
\label{fig:specmap}
\end{figure}

Distributions of the spectral indices on the sky measured at low and high
frequency intervals are shown in Figure \ref{fig:specmap}.
The Galactic plane region has spectral indices of $\alpha\approx-2.5$ -- $-2.0$
and $\beta \approx -1.0$ -- $+1.0$ while the high latitude region with low
foreground contamination shows large fluctuations, especially for $\beta$ map.
It should be noted that the spectral index is a sensitive function of both 
frequency and direction, and that the foreground subtraction of ILC will 
be most successful when the ILC is performed at a region where the foreground 
characteristic is most homogeneous.

\begin{figure*}
\plotone{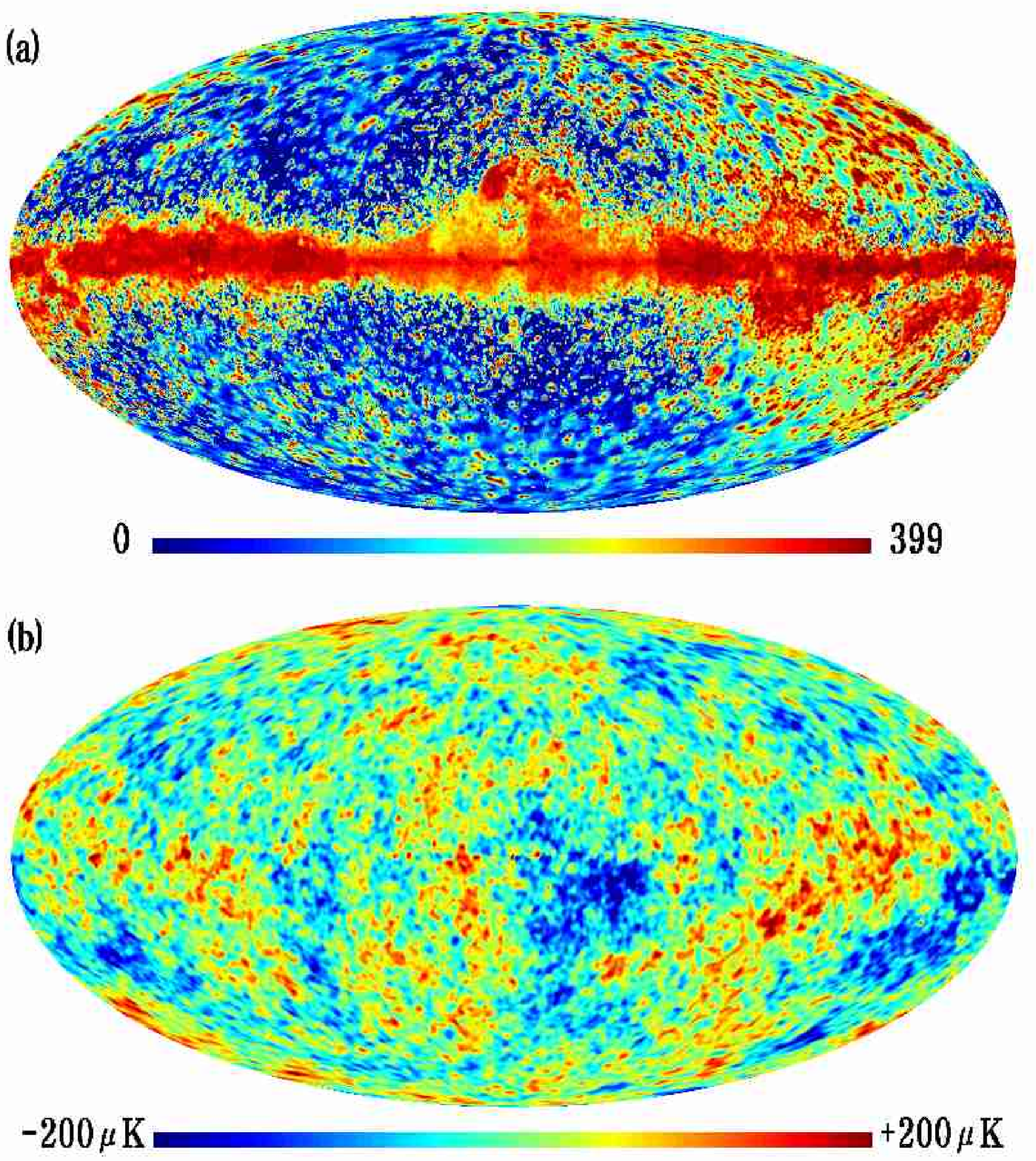}
%\mbox{\epsfig{file=s4.eps,width=88mm,clip=}}
%\resizebox{\textwidth}{!}{\includegraphics{s4.eps}}
\caption{($a$) A group index map showing four hundred groups of pixels
         with similar foreground properties defined by using the foreground 
         spectral index maps at low and high frequency bands 
         (Fig. \ref{fig:specmap}). The group index runs from $0$ to $399$.
         ($b$) The foreground-reduced CMB temperature fluctuation map (SILC400)
         derived from the five-band WMAP maps by applying the ILC method
         which adopts the pixel-groups as defined in ($a$). The SILC400 has 
         been corrected for the residual bias, and the monopole and dipole 
         have been removed from the map. Due to the further smoothing with 
         a Gaussian filter of $\textrm{FWHM}=1\degr$, the map resolution is 
         $1\fdg414$ FWHM. 
         The SILC400 is available at 
         http://newton.kias.re.kr/\~{}parkc/CMB/SILC400.html
         together with its spherical harmonic coefficients $a_{lm}$ up to 
         $l=200$.}
\label{fig:SILC400}
\end{figure*}

For each of the two histograms of spectral indices at the high and low
frequency bands (Fig. \ref{fig:specdistr}$b$), we define twenty spectral
index bins so that each bin contains equal number ($157,300$) of pixels 
with similar foreground spectral properties.
By combining the two sets of spectral-index bins, one at low and the other
at high frequency band, we define four hundred ($20\times 20 = 400$) groups 
of pixels with similar spectral indices.
Because pixels within the same $\alpha$-bin have a wide range of spectral
index $\beta$, the number of pixels contained in each group differs from 
group to group. The resulting group index map is shown in Figure 
\ref{fig:SILC400}$a$. The variation of the spectral index across the sky 
is naturally taken into account in this way.
We note that pixels at the high latitude regions are assigned with a wide 
range of group indices. The ILC will be applied separately for each group 
of pixels with similar foreground spectral properties to obtain 
foreground-reduced CMB temperatures in the next section.

\section{Application to the WMAP Three-Year data}
\label{sec_appl}

\subsection{The ILC Map Derived From the WMAP Three-Year Maps: SILC400}
\label{subsec_applwmap}

\begin{figure*}
\plotone{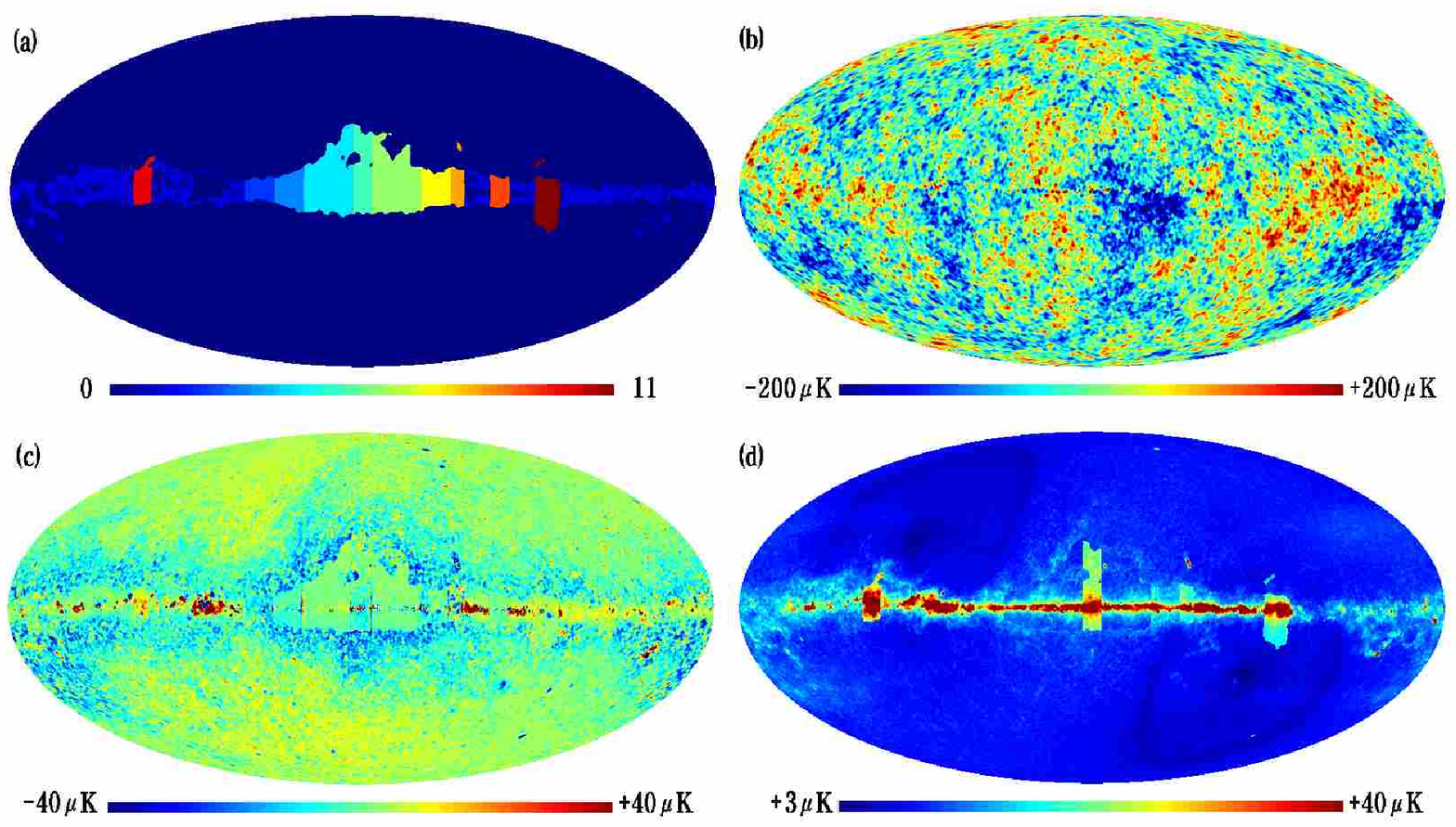}
\caption{($a$) Region definition map defined by the WMAP team (12 regions).
         ($b$) A foreground-reduced CMB map derived with the minimum-variance 
         ILC method (WILC12). The ILC has been applied to 12 regions 
         independently.
         Maps showing ($c$) the average and ($d$) the standard deviation of 
         the residual foreground emissions in the ILC map estimated from two 
         hundred WILC12 simulation maps. The monopole and dipole components 
         of the average map have been removed.}
\label{fig:WILC12}
\end{figure*}

We apply our ILC method to the WMAP three-year 
data\footnote{http://lambda.gsfc.nasa.gov}. The WMAP has one K band 
($22.8$ GHz), one Ka band ($33.0$ GHz), two Q band ($40.7$ GHz), two V band 
($60.8$ GHz), and four W band ($93.5$ GHz) differencing assemblies, with
$0\fdg82$, $0\fdg62$, $0\fdg49$, $0\fdg33$, and $0\fdg21$ FWHM beam widths, 
respectively. The WMAP maps are made in the 
HEALPix\footnote{http://healpix.jpl.nasa.gov} format 
with $\textrm{Nside}=512$ \citep{gor99,gor05}. The total number of pixels 
of a map is $12\times\textrm{Nside}^2 = 3,145,728$.

We use the five band-averaged WMAP maps with $1\degr$ FWHM resolution
that are presented by the WMAP team. The maps have been produced in the 
following way. Each map of each differencing assembly is deconvolved with
the corresponding channel-specific beam and convolved with 
$\textrm{FWHM}=1\degr$ beam. The maps of the same frequency band are averaged 
with noise-weight at each pixel taken into account, and finally the five WMAP 
maps at K, Ka, Q, V, and W bands with $\textrm{FWHM}=1\degr$ resolution 
are obtained.

We obtain a foreground-reduced ILC map (hereafter WILC12) from the three-year 
WMAP data with $1\degr$ FWHM resolution at five frequency bands ($k=5$) 
by using equations (\ref{eq:ilc_temp}), (\ref{eq:w}) and (\ref{eq:st}) 
for each of 12 disjoint regions that are defined by the WMAP team (Figs. 
\ref{fig:WILC12}$a$ and $b$). Our weight coefficients are slightly different 
from those found by \citet{hin06} who compute minimum-variance ILC weights 
by using Lagrange multipliers, but our weight sets give smaller variances.
To estimate the residual bias in the WILC12 map we generate two hundred maps
mimicking the WILC12 map by combining theoretical CMB temperature signal 
with instrument noise and foreground emissions. The three-year version of 
MEM Galactic foreground maps have been used as the contaminating sources.
The average and the standard deviation maps of residual emission in two 
hundred WILC12 simulation maps with respect to the true CMB temperatures 
are shown in Figures \ref{fig:WILC12}$c$ and $d$.
From the WILC12 simulations, we found that ILC applied to the large high 
latitude region that contains pixels with a wide range of foreground spectral 
indices (e.g., the region with index $\mathcal{R}=0$) induces residual biases 
whose levels are different at different positions on the sky. 
As shown in Figure \ref{fig:WILC12}$c$, the residual emission is negative 
near the Galactic plane and is positive at high latitude region. 

\begin{figure}
\mbox{\epsfig{file=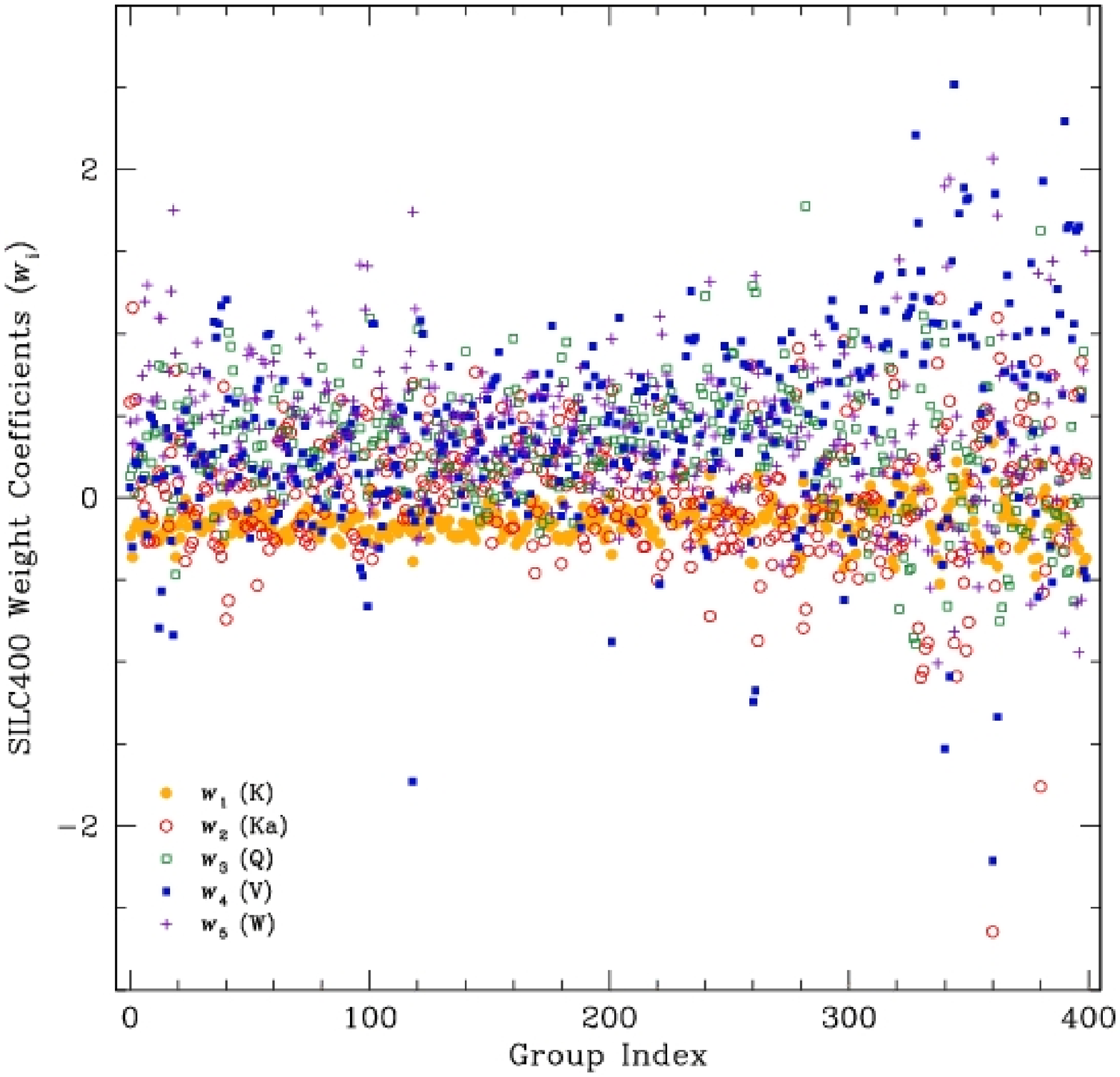,width=88mm,clip=}}
%\resizebox{\textwidth}{!}{\includegraphics{s6.eps}}
\caption{Minimum-variance ILC weights of SILC400 obtained for four hundred
         pixel-groups. The ILC weights are denoted as yellow filled circles 
         for K ($w_1$), red open circles for Ka ($w_2$), green open squares 
         for Q ($w_3$), blue filled squares for V ($w_4$), and violet crosses 
         for W ($w_4$) bands.}
\label{fig:SILC400weight}
\end{figure}

Figure \ref{fig:SILC400}$b$ shows a new foreground-reduced CMB map obtained 
by applying the ILC method described in $\S\ref{sec_ilcmethod}$ with the 
minimum variance constraint, where the combinations are performed for four 
hundred pixel-groups of common spectral properties independently (hereafter 
SILC400). The SILC400 shown here is a map that has been corrected for the 
residual bias based on the ILC simulation as described in $\S\ref{ILC:sim}$, 
and the monopole and dipole have been subtracted from the map. The map has 
been shown after further smoothed by a Gaussian filter of 
$\textrm{FWHM}=1\degr$ to reduce discontinuities between boundaries of pixel 
groups. In the subsequent analysis of large scale modes, however, we use 
the unsmoothed SILC400.

ILC combination weights for 400 groups are plotted in Figure 
\ref{fig:SILC400weight}. The combination weights varies dramatically
from group to group, except for the K-band weights ($w_1$; filled circles)
that are stably distributed around $-0.25$.
Most of the weights in Ka--W bands ($w_2$--$w_5$) are positive for group index 
$\lesssim 250$ corresponding to high latitude regions, while they are 
fluctuating at the Galactic plane regions (group index $\gtrsim 250$).

\begin{figure}
\mbox{\epsfig{file=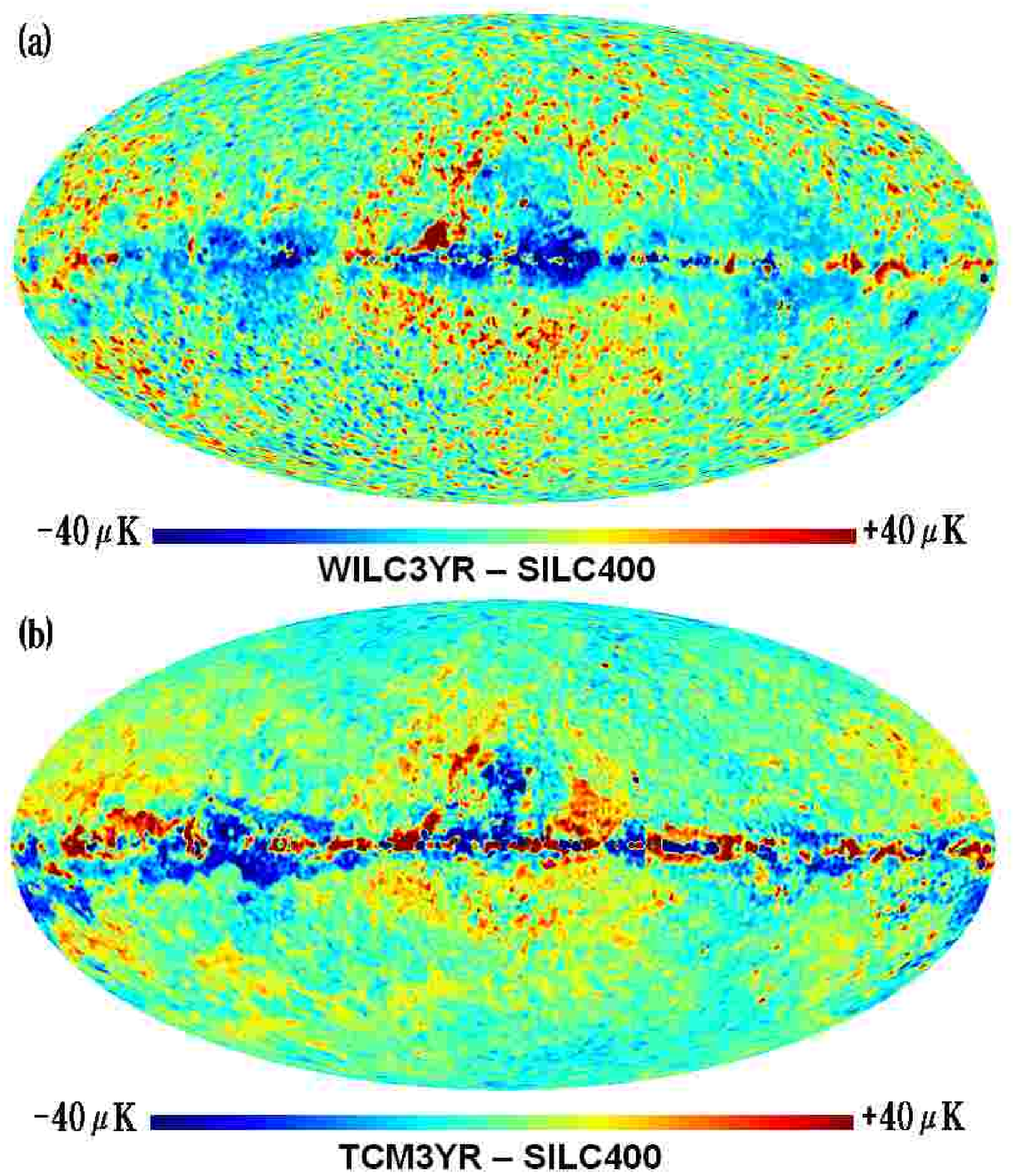,width=88mm,clip=}}
\caption{Comparison of SILC400 with other foreground-reduced CMB maps
         ($a$) WILC3YR and ($b$) TCM3YR. Shown are differences of WILC3YR 
         and TCM3YR from SILC400. Before differencing, all maps are prepared 
         to have $\textrm{FWHM}=1\fdg414$ resolution and the monopole 
         and dipole modes are removed.}
\label{fig:Diff}
\end{figure}

Comparisons of SILC400 with other foreground-reduced maps (WILC3YR and TCM3YR)
are shown as difference maps in Figure \ref{fig:Diff}, where all maps are made
to have $1\fdg414$ FWHM resolution with the monopole and the dipole components 
removed before differencing. 
The WILC3YR is similar to SILC400 within $40~\mu\textrm{K}$ difference,
except for the Galaxy center and cirrus regions.
The cirrus is apparent as red and the Galactic center looks partly blue.
For TCM3YR, its difference from SILC400 shows the region boundaries near 
the Galactic plane and mid-latitude, especially at $l=0\degr$ -- $180\degr$ 
(left part in Fig. \ref{fig:Diff}$b$), where positive and negative temperatures
appear interchangeably. Those region boundaries were defined by \citet{teg03}.

\begin{deluxetable}{ccrrccrr}
\tablewidth{0pt}
%\tablewidth{220pt}
\tabletypesize{\small}
\tablecaption{Low $l$-mode spherical harmonic coefficients of the 
              Minimum-Variance ILC Map SILC400 \label{tab:alm_SILC400}}
\tablecolumns{8}
\tablehead{$l$ & $m$ & Re[$a_{lm}$] & Im[$a_{lm}$] & $l$ & $m$ & Re[$a_{lm}$] & Im[$a_{lm}$]}
\startdata
2 & 0 & $7.51$   & $0.00$     &  5 & 0 & $14.33$  & $0.00$\\
  & 1 & $-1.54$  & $4.82$     &    & 1 & $23.96$  & $1.09$\\
  & 2 & $-18.54$ & $-18.01$   &    & 2 & $-7.75$  & $2.78$ \\
3 & 0 & $-5.25$  & $0.00$     &    & 3 & $19.75$  & $3.58$\\
  & 1 & $-10.74$ & $3.76$     &    & 4 & $-4.21$  & $8.53$\\
  & 2 & $22.54$  & $2.13$     &    & 5 & $9.48$   & $20.26$\\
  & 3 & $-13.52$ & $30.10$    &  6 & 0 & $-1.76$  & $0.00$ \\
4 & 0 & $23.31$  & $0.00$     &    & 1 & $0.78$   & $4.17$\\
  & 1 & $-6.97$  & $9.12$     &    & 2 & $10.39$  & $-2.75$\\
  & 2 & $11.14$  & $4.26$     &    & 3 & $-9.02$  & $-0.28$\\
  & 3 & $7.34$   & $-21.02$   &    & 4 & $13.73$  & $-1.56$     \\
  & 4 & $1.75$   & $-9.68$    &    & 5 & $-5.07$  & $-3.74$  \\
  &   &          &            &    & 6 & $3.64$   & $8.91$ \\[-2mm]
\enddata
\tablecomments{The coefficients are calculated in Galactic coordinates in 
      units of $\mu\textrm{K}$. The $l=2$ coefficients for WILC3YR
      are $a_{20}=11.48$, $a_{21}=-0.05+4.86i$, and $a_{22}=-14.41-18.80i$.
      For TCM3YR, $a_{20}=3.34$, $a_{21}=0.26+4.88i$, and 
      $a_{22}=-14.88-17.26i$.}
\end{deluxetable}

Low $l$-mode spherical harmonic coefficients ($a_{lm}$) of SILC400 are
listed in Table \ref{tab:alm_SILC400} ($l=2$ coefficients for WILC3YR and 
TCM3YR are summarized in the note).
We note that the SILC400 has $a_{20}$ ($=7.51~\mu\textrm{K}$) that is between 
those of TCM3YR ($3.34~\mu\textrm{K}$) and WILC3YR ($11.5~\mu\textrm{K}$), 
and also has slightly different $a_{22}$.
The amplitudes of $a_{20}$ and $a_{22}$ are bigger than $a_{21}$ both in 
SILC400 and the Galactic foreground map (Table \ref{tab:Galfore}).
However, $a_{20}$ and $a_{22}$ have different signs in two maps: 
the quadrupole of CMB and the Galaxy are negatively correlated
(see $\S\ref{sec_discuss}$). This is mainly due to the big cold spot 
in the CMB map at the Galactic center region (see Fig. \ref{fig:SILC400}).

\subsection{Simulations: Test for the ILC Method}
\label{ILC:sim}

As shown in $\S\ref{subsec_applwmap}$, SILC400 differs from foreground-reduced
CMB maps previously made by others.
Because the ILC method conserves the blackbody nature of the CMB signal,
all maps contain exactly the same information for CMB temperature fluctuations
but with different level of foreground residuals.
To quantify the level of the residual foreground in our variance-minimized
ILC map, we have performed two hundred simulations that mimic the WMAP data
and analyzed them in the same way that SILC400 is made.

First, we simulate two hundred WMAP Gaussian CMB signal maps for each 
differencing assembly at each band. The concordance flat $\Lambda\textrm{CDM}$ 
model power spectrum that fits to the WMAP data only, has been used 
\citep{spe06}. For each mock observation, the same $a_{lm}$'s are used 
for all frequency channels. During the map generation, the WMAP beam transfer 
function $B_l$ is used for each differencing assembly, and the instrument 
noise at each pixel is randomly drawn from the Gaussian distribution with 
variance of $\sigma_0^2 / N_{\textrm{obs}}$, where $N_{\textrm{obs}}$ 
is the effective number of observations at each pixel, and $\sigma_0$ 
is the global noise level of the map (Table 1 of \citealt{hin06}).

Secondly, for each differencing assembly realization, we deconvolve the beam
effect, and convolve the map to a common resolution of $1\degr$ Gaussian
FWHM. Differencing assembly maps at the same band are averaged with
weights given by the noise variance at each pixel to produce an average map
at each frequency band.
Finally, the Galactic foreground map (with $1\degr$ FWHM resolution)
at each frequency is co-added to the average WMAP CMB map.
We do not use the Galactic template method to mimic the Galactic foreground
emissions as used in \citet{eri04:ilc}, but use the Galactic foreground
maps derived by MEM from the three-year WMAP data.
Compared to weighted average of template maps, these foreground maps
reasonably describe the Galactic foreground emission even at the Galactic
plane. However, the Galactic foreground maps derived by the MEM are somewhat 
noisy because of the MEM reconstruction noise.
For each simulation, the five foreground-added CMB maps are put into the same
ILC pipeline as used in $\S\ref{subsec_applwmap}$.

\begin{figure}
\mbox{\epsfig{file=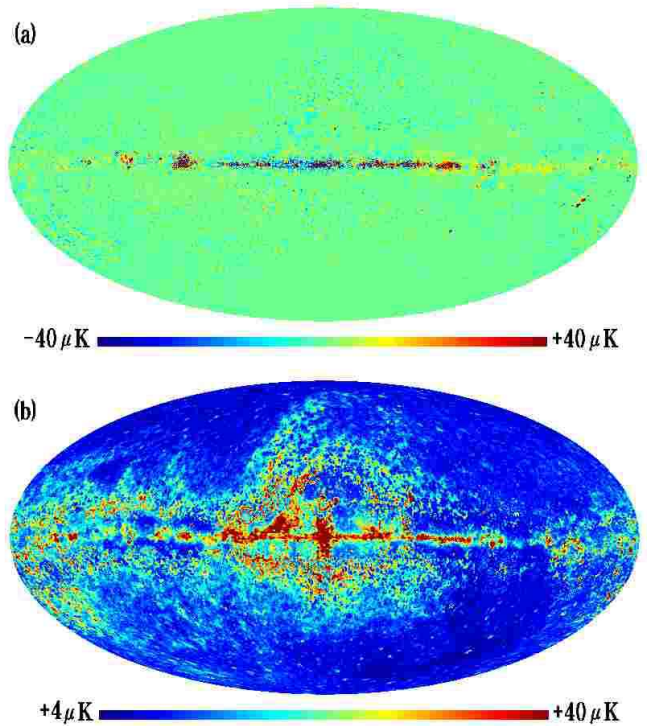,width=88mm,clip=}}
%\resizebox{\textwidth}{!}{\includegraphics{f8.eps}}
\caption{($a$) Average and ($b$) standard deviation maps of the residual
         foreground emissions with respect to true CMB temperatures
         in two hundred SILC400 simulation maps.}
\label{fig:ResILC}
\end{figure}

Figure \ref{fig:ResILC} shows average and standard deviation maps of the 
residual foreground emissions with respect to true CMB temperatures in 
two hundred SILC400 simulation maps. 
It demonstrates that except for the Galactic plane region our ILC method
correctly reconstructs the CMB signal without significant bias due to 
the Galactic foreground contamination unlike the average 
residual-emission map of WILC12 simulations (Fig. \ref{fig:WILC12}$c$). 
However, some residual foreground features with $10$--$30~\mu\textrm{K}$
levels are seen at the high Galactic latitude regions, where the foreground
emission is very weak at high frequency bands (V and W) and the spectral 
index $\beta$ has been measured with large uncertainty due to the MEM 
reconstruction noises (upper right part of Fig. \ref{fig:ResILC}$a$).
The standard deviation map indicates that SILC400 may be contaminated
by the residual foreground at the Galactic plane and the mid-latitude regions.
Compared with the WILC12 case (Fig. \ref{fig:WILC12}$d$), the standard
deviation map has larger amplitudes at high latitude region, because 
pixel-groups used for SILC400 contain much fewer pixels than the 
$\mathcal{R}=0$ region of WILC12.

\begin{figure}
\mbox{\epsfig{file=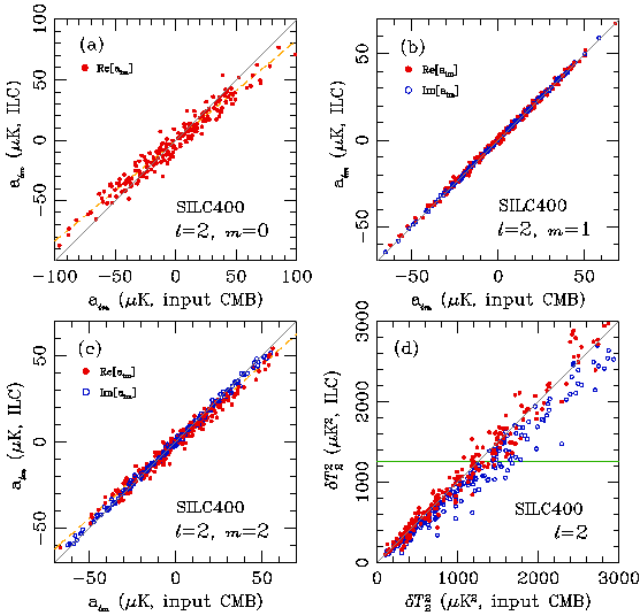,width=88mm,clip=}}
%\resizebox{\textwidth}{!}{\includegraphics{f9.eps}}
\caption{($a$)--($c$) Plots of $l=2$ spherical harmonic coefficients ($a_{lm}$)
         calculated from two hundred SILC400 simulation maps against those 
         from true CMB maps (filled circles for real and the open circles 
         for imaginary components). For real components of even $m$ modes, 
         each plot has been fitted with a straight line (dashed lines).
         ($d$) Plot of quadrupole powers measured from SILC400 simulation
         maps against those from the true CMB maps (open circles).
         Quadrupole powers measured from bias-corrected $a_{lm}$'s of SILC400
         simulations maps are plotted against the true values as filled 
         circles.}
\label{fig:Plot_L2_SILC400}
\end{figure}

Figures \ref{fig:Plot_L2_SILC400}$a$, $b$, and $c$ compare the spherical 
harmonic coefficients of $l=2$ mode calculated from two hundred SILC400 
simulation maps with those of true input CMB maps. To reduce the effect 
of the residual bias, we have subtracted the average map of the residual 
foreground emission from each SILC400 simulation map before calculating 
the $a_{lm}$. The $a_{20}$ and the real component of $a_{22}$ are contaminated 
by the residual Galactic emission in the sense that each plot, when fitted 
with a straight line (dashed lines), gives a slope less than $1$, and some 
scatter is seen in the $a_{20}$ plot. Quadrupole powers from two hundred 
SILC400 simulations against true CMB quadrupole powers are shown as open 
circles in Figure \ref{fig:Plot_L2_SILC400}$d$, where the quadrupole power 
predicted in the concordance $\Lambda\textrm{CDM}$ model is denoted as 
horizontal line ($\delta T_2^2 = 1250$ $\mu\textrm{K}^2$; \citealt{spe06}). 
The SILC400 systematically underestimates the CMB quadrupole powers, but with
much less scatter than LILC simulations (see Fig. 6 of \citealt{eri04:ilc}).

\begin{figure*}
\plottwo{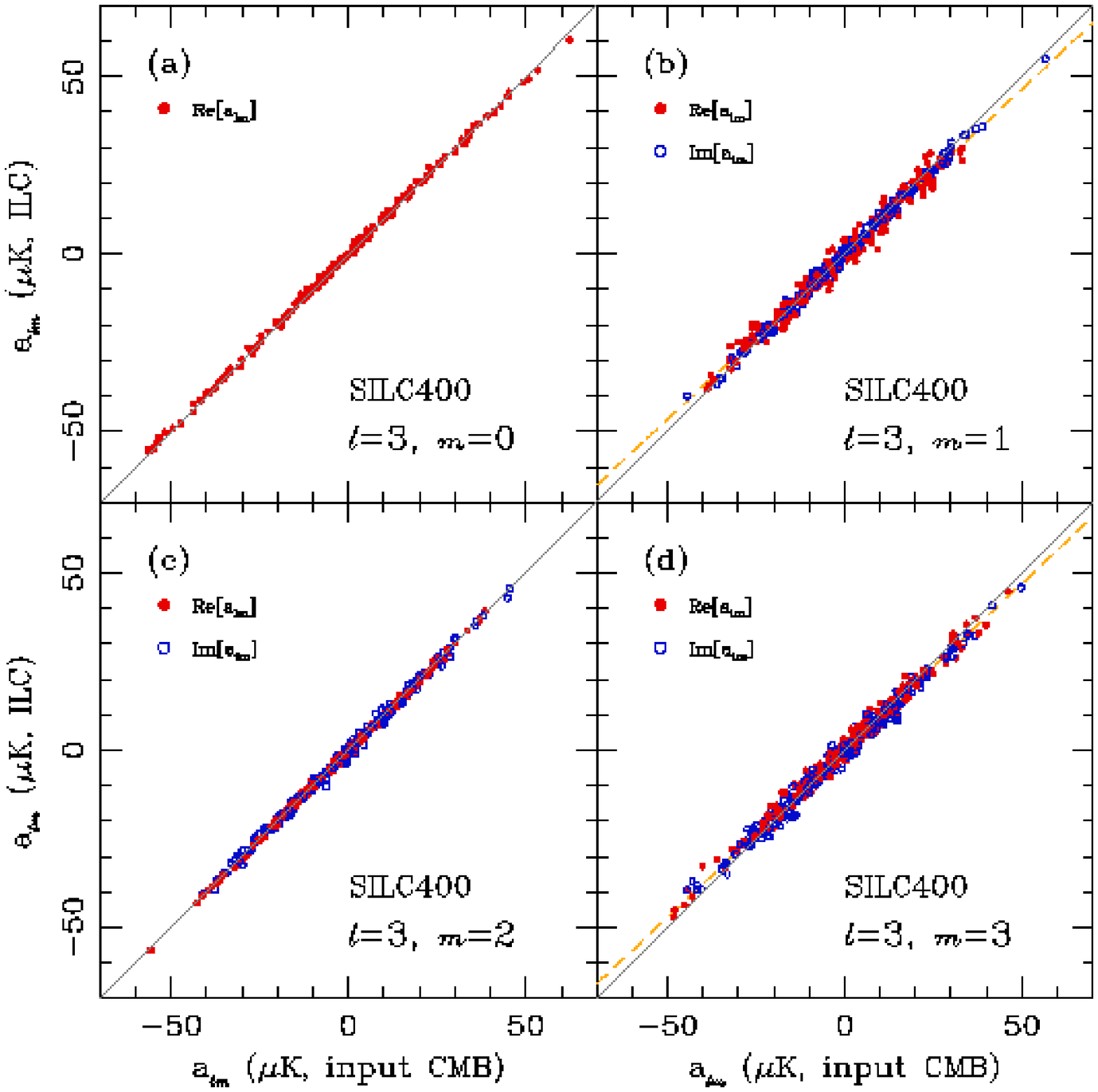}{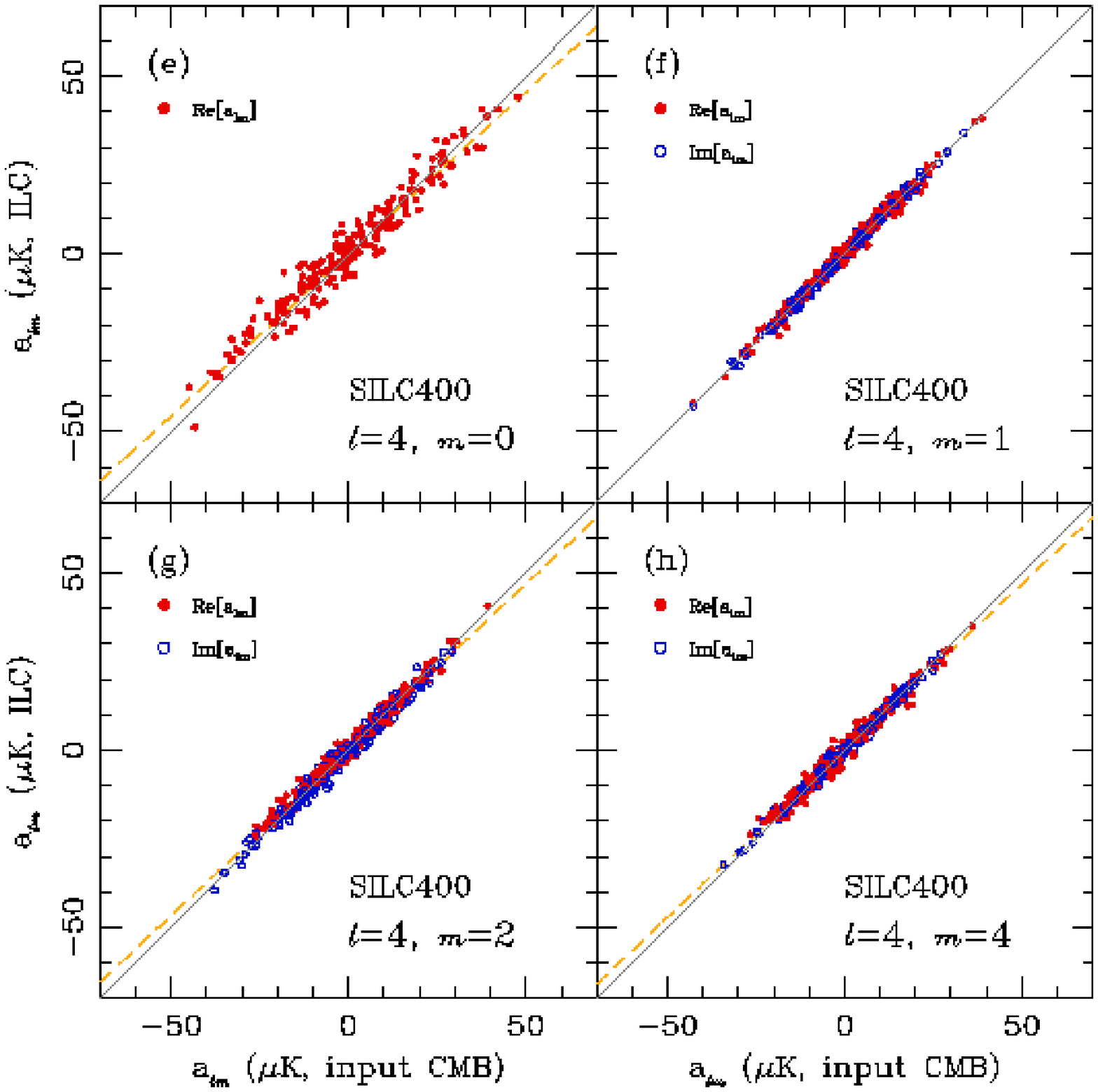}
\caption{Plots of spherical harmonic coefficients ($a_{lm}$'s) estimated 
         from two hundred SILC400 simulation maps against those from true 
         CMB maps for ($a$)--($d$) $l=3$ and ($e$)--($h$) $l=4$ modes.
         The $a_{43}$ plot is omitted because it is very similar to $a_{41}$
         plot.}
\label{fig:Plot_L34_SILC400}
\end{figure*}

Spherical harmonic coefficients for $l=3$ and $4$ of SILC400 simulations
are compared with true values in Figure \ref{fig:Plot_L34_SILC400}.
The $l=3$ modes are excellently reconstructed in the SILC400 simulations,
although plots for real components of $a_{31}$ and $a_{33}$ against true 
values show slight decrease in correlation slopes.
For $l=4$, the real components of $a_{40}$ and $a_{42}$ show similar patterns
to $l=2$ case. The residual foreground emission has induced slope decrease
and scatter in the $a_{40}$ plot (Fig. \ref{fig:Plot_L34_SILC400}$e$).

\begin{figure}
\mbox{\epsfig{file=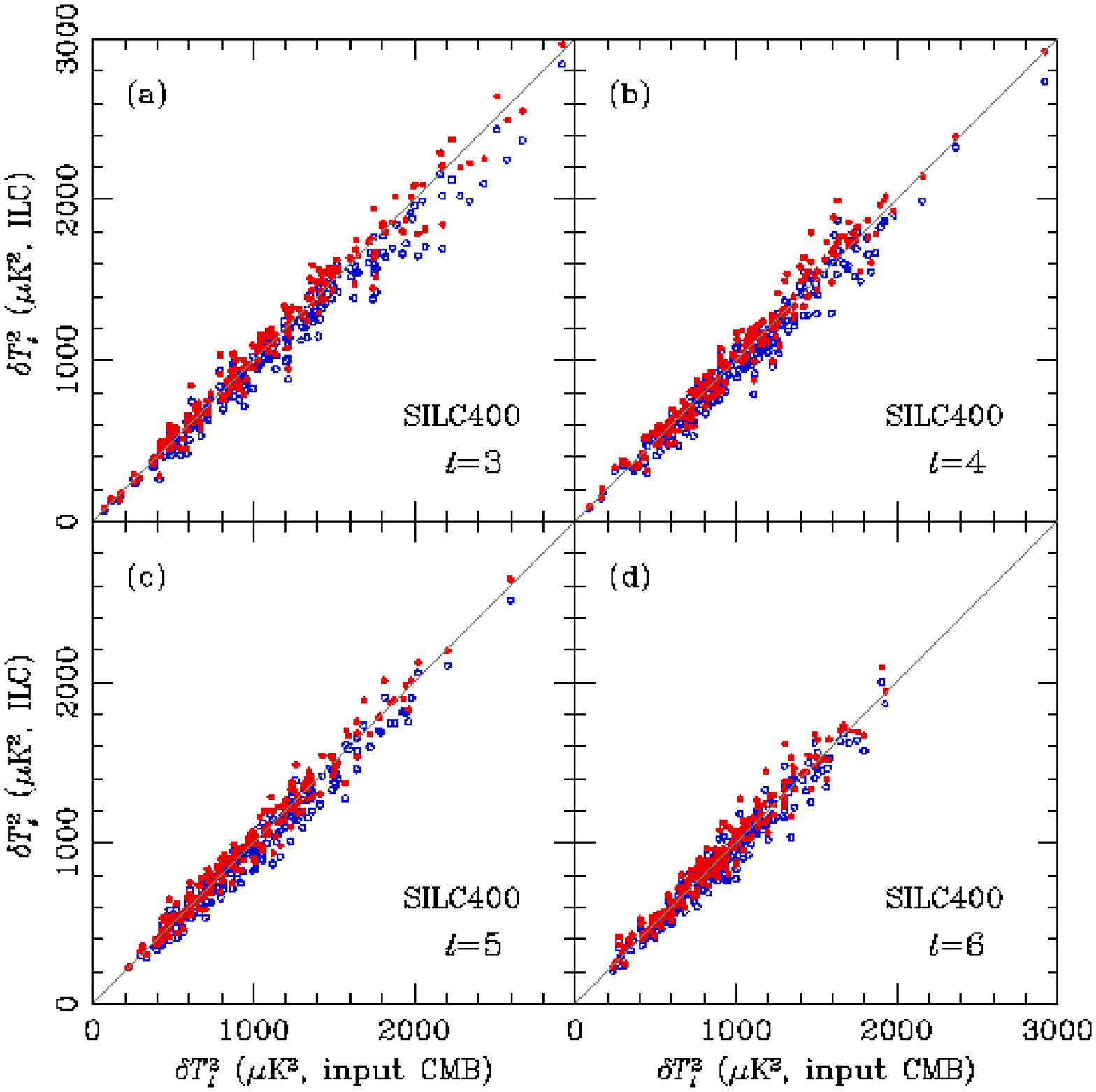,width=88mm,clip=}}
\caption{($a$)--($d$) Plots of low $l$-mode powers ($l=3$--$6$) estimated 
         from two hundred SILC400 simulation maps against those from true 
         CMB maps before (open circles) and after (filled circles) the effect 
         of bias due to the residual Galactic emission on $a_{lm}$'s is 
         removed.}
\label{fig:Plot_powers}
\end{figure}

We can remove the systematic effects of the ILC foreground reduction method
on $a_{lm}$ of the SILC400 simulation maps. We have estimated the true 
$a_{lm}$ up to $l=200$ by applying a simple relation 
$a_{lm} = (a_{lm}^{\textrm{ILC}}-b)/a$, where $a_{lm}^{\textrm{ILC}}$ 
is the spherical harmonic coefficient obtained from each SILC400 simulation 
map from which the average residual bias map of Figure \ref{fig:ResILC}$a$ 
has been subtracted, and $a$ and $b$ denote a correlation slope and an offset 
from the line-fit. Linear fit parameters have been derived from the $a_{lm}$ 
plots, where each slope and offset are obtained for real or imaginary 
components of $a_{lm}$'s, independently.
Real components of $a_{lm}$ for even (odd) $l$ and even (odd) $m$ modes
show the systematic decreases in the slope of the linear relation.
However, in all cases, no large offset is observed in the SILC400 simulations.
The result for $l=2$ is shown as filled circles in Figure
\ref{fig:Plot_L2_SILC400}$d$.
The quadrupole powers of SILC400 simulation maps, which were underestimated
before correction, now have the correct mean amplitudes compared with
the true values. Figure \ref{fig:Plot_powers} shows the low $l$-mode powers of
SILC400 simulations versus true CMB powers for $l=3$--$6$ before (open circles)
and after (filled circles) the bias correction for $a_{lm}$.

\section{Statistics of Large-Scale Modes}
\label{sec_stat}

\begin{deluxetable*}{lccc}
\tablewidth{0pt}
\tabletypesize{\small}
\tablecaption{CMB quadrupole and octopole powers \label{tab:quo_power}}
\tablecolumns{4}
\tablehead{Measurement & $\delta T_2^2$ ($\mu\textrm{K}^2$)  &  $p$ value &
                         $\delta T_3^2$ ($\mu\textrm{K}^2$)}
\startdata
Spergel et al. best-fit $\Lambda$CDM model
                         & $1250$    &  \nodata  &    $1143$ \\
Hinshaw et al. cut sky analysis
                         & $211.0$    &  2.6\%  &    $1041$ \\
WMAP team's ILC map (WILC3YR)
                         & $248.5$    &  3.7\%  &    $1051$ \\
Tegmark et al. cleaned map (TCM)
                         & $201.6$    &  2.3\%  &    $866.1$ \\
de Oliveira-Costa \& Tegmark cleaned map (TCM3YR)
                         & $209.6$    &  2.5\%  &    $1037.8$ \\
Eriksen et al. ILC map (LILC)
                         & $350.6$    &  7.6\% &    $1090$ \\[+2mm]
Minimum-variance ILC map (SILC400)
                         & $244.5_{-203.3}^{+83.8}$ &  4.9\%$^{\dagger}$  
                         & $859.4_{-109.0}^{+61.2}$  \\
SILC400 (bias corrected)
                         & $275.8_{-126.0}^{+94.3}$ &  5.7\%$^{\dagger}$ 
                         & $951.9_{-83.0}^{+63.7}$ \\[-2mm]
\enddata
\tablecomments{$p$ value denotes the probability of finding a lower quadrupole
               than the measured value if the concordance $\Lambda$CDM model
               \citep{spe06} is correct. The $p$ values with $\dagger$ have 
               been estimated from equation (\ref{eq:pvalue}).}
\end{deluxetable*}

Table \ref{tab:quo_power} summarizes the quadrupole and octopole powers
measured from the WMAP data in the previous studies, including our new 
estimates. For SILC400 results, the 68\% confidence limits have been deduced
from the distribution of 
$\delta T_{l}^2 (\textrm{ILC}) - \delta T_{l}^2 (\textrm{CMB})$ in two hundred 
SILC400 simulations. The third column of Table \ref{tab:quo_power} lists 
the probabilities ($p$) of the quadrupole power being as low as measured value
if the concordance $\Lambda$CDM model is correct \citep{deo04}.
However, the $p$ values for SILC400 maps have been estimated from 
\begin{equation}
   \hat{p} = \int d Q_\textrm{obs} p(Q < Q_\textrm{obs}) p(Q_\textrm{obs}),
\label{eq:pvalue}
\end{equation} 
where $Q$ denotes quadrupole power and $p(Q < Q_\textrm{obs})$ is the
probability of finding the quadrupole power as low as the observed value
$Q_\textrm{obs}$. The $p(Q_\textrm{obs})$ is the probability of observing
the quadrupole power $Q_\textrm{obs}$ and is calculated from the 
distribution of quadrupole powers from two hundred SILC400 simulations.
The quadrupole power of SILC400 ($244.5_{-203.3}^{+83.8}~\mu\textrm{K}^2$)
is close to that of WILC3YR ($248.5~\mu\textrm{K}^2$) and is between values 
of TCM3YR ($209.6~\mu\textrm{K}^2$) and LILC ($350.6~\mu\textrm{K}^2$).
Our estimates compare with those of \citet{bie04} who have applied power 
equalization filter to the high latitude WMAP data and obtained 
$\delta T_2^2 = 279 \pm 49$ and $\delta T_3^2 = 870 \pm 98$ from TCM.

Also listed in Table \ref{tab:quo_power} are the quadrupole and octopole
powers from SILC400 after removing the expected biases in $a_{lm}$
due to the minimum variance ILC method.
Our bias-corrected quadrupole power ($275.8_{-126.0}^{+94.3}~\mu\textrm{K}^2$) 
is higher than the WMAP team's measurement \citep{hin06}, but they are still 
consistent with each other. The probability of observing such a low quadrupole
power in the concordance $\Lambda\textrm{CDM}$ universe is as high as 
$p = 5.7\%$. 

\begin{figure}
\mbox{\epsfig{file=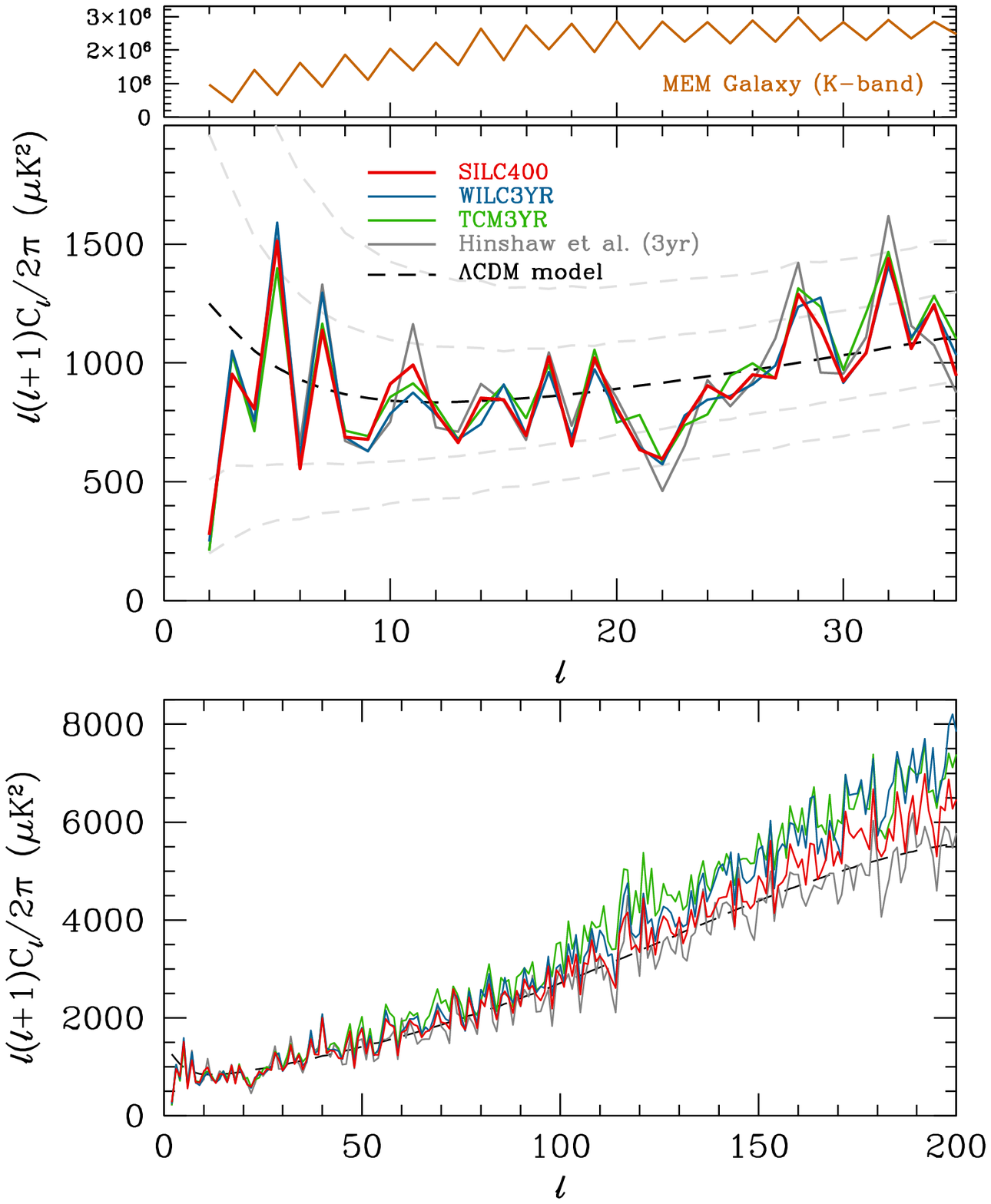,width=88mm,clip=}}
%\resizebox{\textwidth}{!}{\includegraphics{f12.eps}}
\caption{Power spectra measured from SILC400 (thick red), WILC3YR (blue),
         and TCM3YR (green curve). The power spectrum of K-band MEM 
         Galactic foreground map has also been shown for comparison 
         (brown curve). Beam effects are deconvolved for the correct 
         amplitudes of power spectra.
         The WMAP power spectrum measured by \citet{hin06} (grey curve), and
         the $\Lambda\textrm{CDM}$ model of \citet{spe06} (dashed curve)
         are shown for comparison with 68\% and 95\% confidence limits
         (grey dashed curves).}
\label{fig:Plot_pscompare}
\end{figure}
Figure \ref{fig:Plot_pscompare} compares SILC400 power spectrum with other 
measurements up to $l=35$ (top) and $l=200$ (bottom).
Also shown in the top panel is the power spectrum of K-band MEM Galactic
foreground map whose zigzag pattern at $l=2$--$8$ is very similar to that 
of the CMB power spectra. For WILC3YR, TCM3YR, and SILC400, we have deconvolved
the beam effects to obtain correct amplitudes of power spectrum.
The WMAP team's power spectrum measurement from the cut-sky analysis 
and the best-fit $\Lambda\textrm{CDM}$ power spectrum are also shown 
for comparison with 68\% and 95\% confidence limits \citep{hin06,spe06}.
The quadrupole power of the $\Lambda$CDM model is $1250~\mu\textrm{K}^2$
with 68\% and 95\% confidence intervals, $510$--$1961~\mu\textrm{K}^2$ 
and $198$--$3311~\mu\textrm{K}^2$, respectively (from the ten thousand 
realizations).

Our new estimate of power spectrum from SILC400 (red curve) is very similar
to others up to $l \approx 30$. Due to the instrument and reconstruction 
noises, all ILC maps give angular power spectra with higher amplitudes 
with increasing $l$, compared with the $\Lambda$CDM model.
As shown in Figure \ref{fig:Plot_pscompare} (bottom), the angular power 
spectrum of the SILC400 has lower amplitude than those of WILC3YR and TCM3YR 
up to $l=200$, which indicates that our ILC method better satisfies the 
minimum variance constraint.

It has been reported that both the quadrupole and the octopole of the WMAP 
map appear planar, with most of their hot and cold spots located on a single
plane in the sky, and the two planes appear roughly aligned \citep{deo04}.
A simple way to quantify a preferred axis for arbitrary multipoles is to find 
the axis $\hat{\mbox{\boldmath $n$}}$ around which the angular momentum 
dispersion
\begin{equation}
   \left< \psi|(\hat{\mbox{\boldmath $n$}}\cdot{\bf L})^2|\psi\right>
   = \sum_m m^2 |a_{lm}(\hat{\mbox{\boldmath $n$}})|^2
\label{equ:angmom}
\end{equation}
is maximized \citep{deo04}, where the CMB map is considered as a wave function 
$\Delta T (\hat{\mbox{\boldmath $n$}}) \equiv \psi (\hat{\mbox{\boldmath 
$n$}})$. Here $a_{lm}(\hat{\mbox{\boldmath $n$}})$ denotes the spherical 
harmonic coefficients of the CMB map in a rotated coordinate system with 
its $z$-axis in the $\hat{\mbox{\boldmath $n$}}$ direction.
For each $l$ we find a unit vector $(\hat{\mbox{\boldmath $n$}}_l)$ that 
maximizes the angular momentum dispersion.
We start to evaluate equation (\ref{equ:angmom}) for all the unit vectors
corresponding to HEALPix pixel centers at resolution $\textrm{Nside}=32$,
perform the same operation only for pixels at higher resolution around 
the unit vector found in the previous step, and finally obtain a unit vector 
that maximizes equation (\ref{equ:angmom}) at resolution $\textrm{Nside}=4096$.

Table \ref{tab:quo_dir} lists directions of the quadrupole and the octopole 
and separations between two poles measured from the WMAP data. 
The 68\% confidence limits for the measured angular separation $\theta_{23}$ 
of SILC400 have been estimated from the distribution of 
$\theta_{23}(\textrm{ILC}) - \theta_{23}(\textrm{CMB})$ in two hundred SILC400 
simulations. Since the quantity $|\hat{\mbox{\boldmath $n$}}_2 \cdot
\hat{\mbox{\boldmath $n$}}_3|$ has a uniform distribution on the unit interval 
$\left[0,1\right]$, the probability of finding an angular separation smaller 
than the measured separation is simply given by 
$p=1-|\hat{\mbox{\boldmath $n$}}_2 \cdot \hat{\mbox{\boldmath $n$}}_3|$ 
(last column in Table \ref{tab:quo_dir}). 
However, the $p$ values for SILC400 maps have been calculated from equation 
(\ref{eq:pvalue}) with $Q$ replaced with $\theta_{23}$. The bias-corrected 
SILC400 has an angular separation ($\theta_{23}=11\fdg8_{-8\fdg0}^{+6\fdg4}$) 
that is somewhat larger than those from TCM, LILC, and WILC3YR but is very 
similar to that of TCM3YR. We find that the octopole directions
($\hat{\mbox{\boldmath $n$}}_3$) are more stable than the quadrupole directions
($\hat{\mbox{\boldmath $n$}}_2$). Our result for SILC400 confirms the previous 
results that the quadrupole and octopole directions are aligned 
(\citealt{teg03,eri04:ilc,bie04,cop04,cop06,sch04}), with $p=4.3$\%. 

\begin{deluxetable*}{lcccccc}
\tablewidth{0pt}
\tabletypesize{\small}
\tablecaption{Directions of the Quadrupole and the Octopole\label{tab:quo_dir}}
\tablecolumns{7}
\tablehead{Maps & $l_2$ &  $b_2$ & $l_3$ & $b_3$ & $\theta_{23}$ & $p$ value}
\startdata
WMAP team's ILC map (WILC3YR) 
 & $235\fdg23$ & $68\fdg42$ & $236\fdg94$ & $62\fdg61$ & $5\fdg85$ & 0.52\% \\
Tegmark et al. clean map (TCM)  
 & $257\fdg58$ & $58\fdg82$ & $238\fdg37$ & $62\fdg04$ & $9\fdg96$ & 1.51\% \\
TCM3YR (de Oliveira-Costa \& Tegmark 2006)  
 & $224\fdg04$ & $76\fdg56$ & $236\fdg75$ & $63\fdg98$ & $13\fdg23$ & 2.65\% \\
Eriksen et al. ILC map (LILC)  
 & $247\fdg52$ & $61\fdg90$ & $232\fdg84$ & $63\fdg41$ &  $6\fdg89$ & 0.72\% \\[2mm]
Minimum-variance ILC map (SILC400)    
 & $240\fdg78$ & $75\fdg24$ & $234\fdg47$ & $62\fdg26$ & $13\fdg16_{-6\fdg79}^{+7\fdg62}$ & 5.36\%$^\dagger$ \\
SILC400 (bias corrected)   
 & $245\fdg54$ & $74\fdg38$ & $234\fdg58$ & $63\fdg25$ & $11\fdg77_{-8\fdg03}^{+6\fdg40}$ & 4.29\%$^\dagger$ \\[-2mm]
\enddata
\tablecomments{$\theta_{23}$ is the angular separation defined as
        $\theta_{23}=\cos^{-1} (\hat{\mbox{\boldmath $n$}}_2 \cdot 
        \hat{\mbox{\boldmath $n$}}_3)$, where $\hat{\mbox{\boldmath $n$}}_2
        =(l_2,b_2)$ and $\hat{\mbox{\boldmath $n$}}_3=(l_3,b_3)$.
        The $p$ values with $\dagger$ have been estimated from equation 
        (\ref{eq:pvalue}) with $Q$ replaced by $\theta_{23}$.}
\end{deluxetable*}

\begin{figure}
\mbox{\epsfig{file=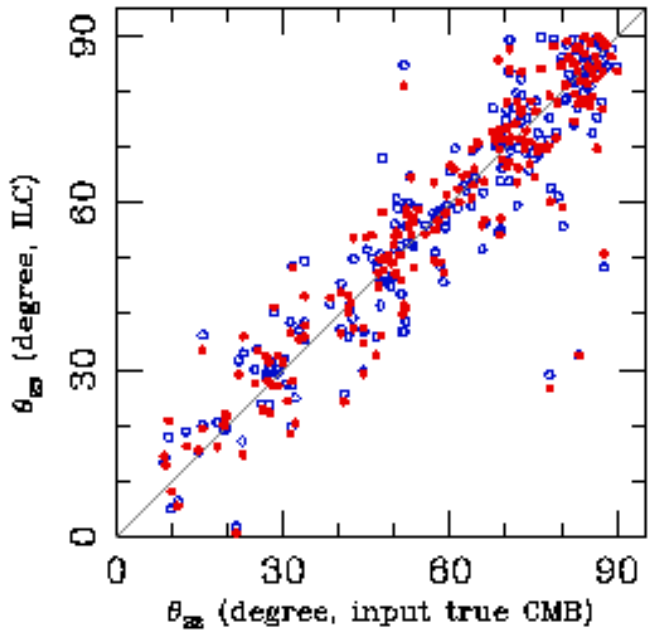,width=88mm,clip=}}
\caption{Plots of angular separation ($\theta_{23}$) between the quadrupole
         and the octopole directions measured from two hundred SILC400
         simulation maps against those from true CMB maps,
         before (open circles) and after (filled circles) the bias correction.}
\label{fig:Plot_pol23}
\end{figure}

Figure \ref{fig:Plot_pol23} compares angular separations between quadrupole 
and octopole directions of two hundred SILC400 simulation maps with the true 
values, and demonstrates that SILC400 can reconstruct the correct multipole 
directions of low $l$-modes.

\begin{figure}
%\plotone{s14.eps}
\mbox{\epsfig{file=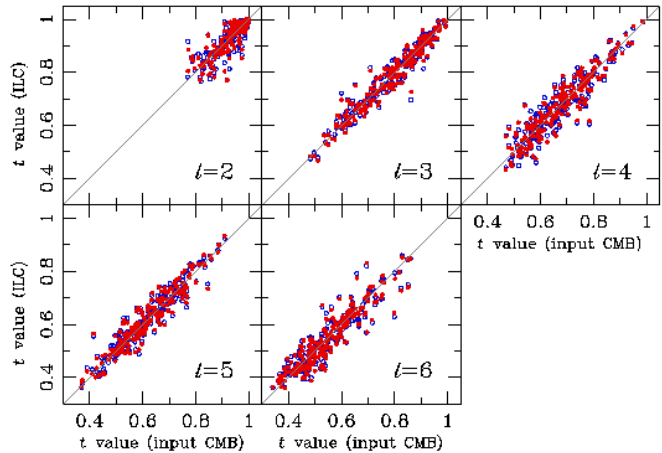,width=88mm,clip=}}
%\resizebox{\textwidth}{!}{\includegraphics{s14.eps}}
\caption{Plots of $t$-statistic for $l=2$--$6$ modes measured from two hundred 
         SILC400 simulation maps against those from true CMB maps, before 
         (open circles) and after (filled circles) the bias correction.}
\label{fig:planar}
\end{figure}

We also measure the degree of planarity of low $l$ modes by calculating
the $t$ statistic defined by \citet{deo04} as
\begin{equation}
   t = \max_{\hat{\mbox{\boldmath $n$}}} {{|a_{l-l}|^2 + |a_{ll}|^2}
          \over {\sum_{m=-l}^{+l} |a_{lm}|^2 }}.
\label{equ:t_stat}
\end{equation}
The $t$ statistic measures the maximal percentage of $l$-mode power
that contributes to $|m|=l$, and we obtain $t$ value for each $l$
by finding a direction $\hat{\mbox{\boldmath $n$}}$ that maximizes equation
(\ref{equ:t_stat}). The maximization is performed over pixels of a map
in the similar way as used in finding the quadrupole and octopole directions.
The performance of our ILC method in reconstructing the true $t$-statistic
is shown in Figure \ref{fig:planar}, where $t$ values of SILC400 simulation
maps against those from true CMB maps are plotted for $l=2$--$6$ modes.

The $t$ values measured from the foreground-cleaned maps are summarized 
in Table \ref{tab:planar}. Numbers in the parentheses represent the number 
of occurrences that have larger $t$ values than the measured value among 
the two hundred SILC400 simulations. The 68\% confidence limits for $l=2$ 
and $3$ are obtained from the distributions of 
$t(\textrm{ILC})-t(\textrm{CMB})$ in the SILC400 simulations.
The $t$ value for $l=3$ mode of SILC400 (bias corrected; 
$t=0.91_{-0.03}^{+0.02}$) is very similar to the previously measured values,
and implies that the octopole has high planarity.
However, the planarity is not statistically significant because the 
probability that SILC400 simulation map has $t$ value larger than the 
measured $t$ value is 18.5\%: thirty seven among two hundred SILC400 
simulations. The probability of having the $t$ value for the quadrupole mode 
as high as the measured value of SILC400 ($t=0.98_{-0.02}^{+0.02}$) is still 
higher (98/200). As shown in Figure \ref{fig:planar}, most of the CMB 
quadrupole modes have $t>0.8$.

\begin{deluxetable*}{llllll}
\tablewidth{0pt}
\tabletypesize{\small}
\tablecaption{$t$ Values of Low Spherical Harmonic Modes\label{tab:planar}}
\tablecolumns{7}
\tablehead{Maps & $l=2$ &  $l=3$ & $l=4$ & $l=5$  &  $l=6$}
\startdata
WMAP team's ILC map (WILC3YR) 
   & $0.930$ & $0.917$ & $0.648$ & $0.375$ & $0.798$ \\
Tegmark et al. clean map (TCM)
   & $0.957$ & $0.942$ & $0.588$ & $0.372$ & $0.783$ \\
TCM3YR (de Oliveira-Costa \& Tegmark 2006) 
   & $0.995$ & $0.877$ & $0.679$ & $0.358$ & $0.749$ \\
Eriksen et al. ILC map (LILC)
   & $0.913$ & $0.934$ & $0.587$ & $0.374$ & $0.806$ \\[2mm]
Minimum-variance ILC map (SILC400)
   & $0.991_{-0.021}^{+0.017}$ (61) & $0.901_{-0.027}^{+0.029}$ (38) & $0.523$ (180) & $0.377$ (199) & $0.835$ (3) \\
SILC400 (bias corrected)
   & $0.977_{-0.024}^{+0.016}$ (98) & $0.908_{-0.028}^{+0.024}$ (37) & $0.530$ (177) & $0.382$ (199) & $0.826$ (4) \\[-2mm]
\enddata
\tablecomments{Numbers in the parentheses represent the number of occurrences 
       that give larger $t$ values than the measured value among the two 
       hundred SILC400 simulations.}
\end{deluxetable*}

As pointed by \citet{eri04:ilc}, the $l=5$ and $l=6$ modes are very peculiar
in their symmetry properties. Only one case out of the two hundred SILC400 
simulations has $t$ values larger than $t=0.38$ for $l=5$, and only four have 
$t$ values larger than $t=0.83$ for $l=6$,
which indicates that the distribution of temperature fluctuation of $l=5$ mode 
is highly symmetric, and the $l=6$ mode is planar together with the $l=2$ and 
$l=3$ modes.

\begin{figure}
\mbox{\epsfig{file=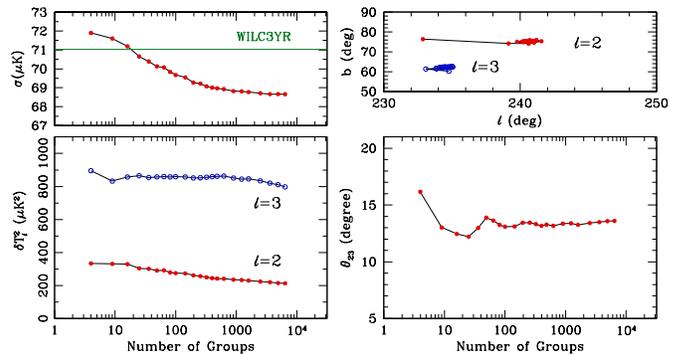,width=88mm,clip=}}
%\resizebox{\textwidth}{!}{\includegraphics{s15.eps}}
%\plotone{s15.eps}
\caption{Standard deviations ($\sigma$; top-left), the quadrupole and
         the octopole powers ($\delta T_{l}^2$; bottom-left), and angular 
         separations between the quadrupole and the octopole ($\theta_{23}$; 
         bottom-right panel) measured from ILC maps made for various number 
         of pixel-groups from $4$ to $6400$.
         The top-right panel shows the variation of the quadrupole
         (filled circles) and the octopole (open circles) directions
         in Galactic coordinates. The horizontal line in the top-left panel
         denotes the standard deviation of the WILC3YR map.}
\label{fig:stat_compare}
\end{figure}

We also investigate whether the measured statistics depend on the number 
of pixel-groups and the smoothing scale. First, we measure standard deviation, 
quadrupole and octopole powers, and angular separation between the two 
multipoles from the ILC maps made by fixing the resolution to 
$1\degr~\textrm{FWHM}$ while varying the number of pixel-groups 
in the group-index map from $4$ ($2\times 2$) to $6400$ ($80\times 80$).
The result is shown in Figure \ref{fig:stat_compare}.
The standard deviation and quadrupole power decrease with increasing number
of pixel-groups, while the octopole power is very stable.
If the number of pixel-groups is larger than $100$, all statistical quantities
become stable. However, when the number of pixel-groups is small,
a wide range of foreground spectral indices is allowed in a common
group, and the ILC results in higher values of standard deviation, quadrupole
power, and larger angular separation $\theta_{23}$.
We note that $\theta_{23}$ becomes large because the quadrupole direction
is unstable in this case (filled circles in the top-right panel of
Fig. \ref{fig:stat_compare}).

\begin{figure}
\mbox{\epsfig{file=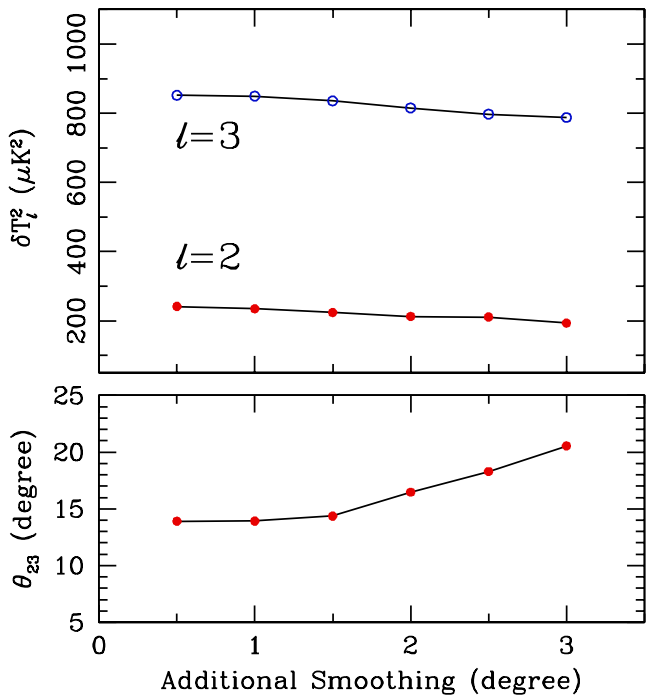,width=88mm,clip=}}
\caption{Quadrupole and octopole powers ($\delta T_{l}^2$) and angular 
         separations between the quadrupole and the octopole ($\theta_{23}$) 
         measured from the ILC maps produced from the WMAP data and the 
         Galactic foreground maps with various angular resolution. 
         Each WMAP or Galaxy map with $\textrm{FWHM}=1\degr$ has been further 
         smoothed with a smoothing filter of $0\fdg5$ to $3\fdg0$ FWHM 
         with steps of $0\fdg5$. Total resolution of ILC maps varies 
         from $1\fdg12$ to $3\fdg16$ FWHM.}
\label{fig:stat_compare_vs}
\end{figure}

Secondly, we measure the large scale mode statistics from the ILC maps 
produced from the WMAP data and the Galactic foreground maps with various 
angular resolution. Each WMAP map or MEM-derived Galactic foreground map,
which has originally $1\degr$ FWHM resolution, has been further smoothed
with a Gaussian filter of $\textrm{FWHM}=0\fdg5$--$3\fdg0$ with steps of 
$0\fdg5$ (the total resolution varies from $1\fdg12$ to $3\fdg16$ FWHM).
The smoothed maps are used to calculate spectral indices for pixel-group
definition and for subsequent application to the ILC pipeline.
In all cases, the number of pixel groups has been fixed to 400.
As shown in Figure \ref{fig:stat_compare_vs}, our ILC method gives stable
and consistent amplitudes of the quadrupole and the octopole powers,
although the angular separation $\theta_{23}$ tends to increase with 
increasing smoothing scale.

\section{Discussion}
\label{sec_discuss}

In this paper, we have derived a new foreground-reduced CMB map by applying 
a simple internal linear combination method to the WMAP three-year data. 
Rather than using the disjoint sky regions as in WILC3YR, TCM, and LILC, 
we have defined a new group index map composed of four hundred pixel-groups 
that contain pixels with similar foreground spectral properties, and obtained 
a CMB map with foreground emission effectively reduced (SILC400).

Two hundred ILC simulations show that the residual foreground emission 
in the SILC400 is very small in amplitude, which is at $10$--$30~\mu\textrm{K}$
levels at high Galactic latitude regions (Fig. \ref{fig:ResILC}$a$).
On high latitude regions where the foreground emission is very weak at high 
frequency bands (V and W), the spectral index is inaccurately measured due 
to the MEM reconstruction noise in the Galactic foreground maps. 
If ILC regions are defined based on the inaccurate spectral index information, 
the performance of foreground removal with the ILC method becomes low. 
To reduce such noises, we have smoothed the foreground maps with a Gaussian 
filter whose width varies depending on the signal-to-noise level of intensities
in the MEM Galaxy maps.

Such residual foreground features do not affect significantly the statistics 
of large-scale modes of CMB anisotropy. For example, when the average residual 
emission map (Fig. \ref{fig:ResILC}$a$) is added to or subtracted from the 
SILC400 map, the estimated quadrupole power and angular separation are 
$\delta T_2^2=288~\mu\textrm{K}^2$ and $\theta_{23}=13\degr$ for the addition 
map and $\delta T_2^2=264~\mu\textrm{K}^2$ and $\theta_{23}=11\degr$
for the subtraction map. Both are within the $68\%$ confidence limits
of the SILC400 results. This implies that the residual foreground emission
in our analysis map is not statistically important to the large-scale modes 
of CMB anisotropy.

We have also shown that the SILC400 recovers the true low $l$-mode powers,
angular separation between the quadrupole and the octopole, and the planarity
parameter (Figs. \ref{fig:Plot_L2_SILC400}--\ref{fig:Plot_powers},
\ref{fig:Plot_pol23}, and \ref{fig:planar}). 
Our SILC400 correctly reconstructs the spherical harmonic coefficients with 
minimal biases. Our ILC method is insensitive to the number of pixel-groups 
of common spectral index and on the smoothing scale as demonstrated in
Figures \ref{fig:stat_compare} and \ref{fig:stat_compare_vs}.

The quadrupole and octopole powers measured from SILC400 are 
$244_{-203}^{+84}~\mu\textrm{K}^2$ and $859_{-109}^{+61}~\mu\textrm{K}^2$
(68\% C.L.), respectively. According to the SILC400 simulations, the minimum 
variance ILC method tends to underestimate the low $l$-mode powers.
Removing the effect of bias due to the residual Galactic emission on the 
spherical harmonic coefficients of the SILC400, we obtain 
$\delta T_2^2 = 276_{-126}^{+94}~\mu\textrm{K}^2$ and 
$\delta T_3^2 = 952_{-83}^{+64}~\mu\textrm{K}^2$. We confirm that the CMB 
quadrupole power is still lower than the theoretical value of the concordance 
$\Lambda$CDM model, with $p = 5.7\%$. Our estimate is consistent with 
that of WILC3YR, and is located between those measured from the TCM3YR and 
LILC.

The quadrupole power of the SILC400 is consistent with the previous
measurements from the high latitude part of CMB data. \citet{efs04} 
applied the maximum likelihood analysis method to the foreground-reduced 
CMB maps, and measured the quadrupole powers, $223~\mu\textrm{K}^2$ 
for WILC1YR and $250~\mu\textrm{K}^2$ for TCM for the Kp2 sky coverage. 
\citet{bie04} applied the power equalization filter to the WMAP data with 
Kp2 mask, and obtained quadrupole powers $248\pm45~\mu\textrm{K}^2$ 
for WILC1YR and $279\pm49~\mu\textrm{K}^2$ for TCM.

The angular separation between the quadrupole and the octopole and the their 
planar properties have also been investigated. We have confirmed that 
the quadrupole and the octopole are aligned with high planarity.
The probabilities of observing such anomalies from the bias-corrected SILC400
map are 4.3\% for angular separation $\theta_{23}$ and over 18\% for planarity 
parameters (see Tables \ref{tab:quo_dir} and \ref{tab:planar}).
The observed angular separation is marginally statistically significant.

As observed in Figure \ref{fig:Plot_pscompare}, the power spectra of the CMB 
and the Galactic emission maps show zigzag patterns at $l=2$--$8$, with the 
zigzag directions opposite to each other, which strongly implies that 
the large scale CMB signal estimated from SILC400 and other analyses 
may be contaminated by the Galactic foreground emission and is anti-correlated 
with the Galactic signal (see also Tables \ref{tab:Galfore} and 
\ref{tab:alm_SILC400}). First, we have investigated whether the observed 
zigzag pattern at $l=2$--$8$ in the CMB power spectrum is statistically 
significant or not. Among the 200 true input CMB signal maps, we have found 
16 cases with the CMB zigzag pattern, 10 cases with the Galaxy zigzag 
direction, and totally 26 cases (13\%). 

Secondly, we have measured cross-correlation at zero lag
$C(0) \equiv \sum_{i} \Delta T_{\textrm{S}}(\hat{\mbox{\boldmath $n$}}_i) 
T_{\textrm{G}} (\hat{\mbox{\boldmath $n$}}_i)/N_{p}$ between the SILC400 
(bias-corrected) containing only $l=2$--$8$ modes and the K-band MEM Galactic 
foreground map smoothed by the $\textrm{FWHM}=7\degr$ Gaussian filter. 
Both maps have been degraded to $\textrm{Nside}=32$ before cross-correlation 
measurement. The $C(0)$ values estimated from pixels at $|b| \le 20\degr$ are
$-0.008$ and $-0.020~\textrm{mK}^2$ for northern and southern hemispheres, 
respectively. We see that the CMB signal on the southern hemisphere 
has strong anti-correlation with the Galactic emission. 
We have calculated the cross-correlation functions between the 200 true input
CMB signal maps and the Galactic emission map. The median values together 
with 68\% confidence limits for $C(0)$ are $0.002_{-0.025}^{+0.024}$ and 
$0.000_{-0.023}^{+0.024}$ for $|b| \le 20\degr$ at the northern and
southern hemispheres, respectively.
The estimated probabilities that the large scale modes of true input 
CMB signal map have anti-correlation with the Galactic emission map 
as low as the measured values by chance are 36.5\% and 19.5\% 
for northern and southern hemispheres, respectively. 
The large scale modes of SILC400 on the southern hemisphere also shows 
anti-correlation with the Galactic emission even at $|b|>20\degr$.
However, whether such anti-correlation occurs due to the residual 
Galactic emission or by chance is not clear and further investigation 
is needed.

In this study, to measure the spectral indices of the foreground,
we have used the MEM-derived Galactic foreground maps that contain 
some level of reconstruction errors and fail to model the Galaxy
on the Galactic plane region.
For the best reconstruction of CMB anisotropy through the ILC method,
it is essential to use the Galactic emission maps that model
the Galaxy realistically for precise measurement of the foreground spectral
indices.

\begin{acknowledgements}
We acknowledge use of the Legacy Archive for Microwave
Background Data Analysis (LAMBDA) and use of the HEALPix software 
for deriving the results in this paper. We also acknowledge Max Tegmark 
for providing us with a software that calculates spherical harmonic 
coefficients in an arbitrarily rotated system.
This work was supported by the Korea Science and Engineering Foundation 
(KOSEF) through the Astrophysical Research Center for the Structure and 
Evolution of the Cosmos (ARCSEC) and through the grant R01-2004-000-10520-0.
JRG acknowledges NSF Grant AST04-06713. 
\end{acknowledgements}


\begin{thebibliography}{}
\bibitem[Abramo et al.(2006)]{abr06}
   Abramo, L.R., Bernui, A., Ferreira, I.S., Villela, T., \& Wuensche, C.A.
   2006, preprint (astro-ph/0604346)
\bibitem[Bennett et al.(2003a)]{ben03:basic}
   Bennett, C.L., et al. 2003a, \apjs, 148, 1
\bibitem[Bennett et al.(2003b)]{ben03:galaxy}
   Bennett, C.L., et al. 2003b, \apjs, 148, 97
\bibitem[Bielewicz et al.(2005)]{bie05}
   Bielewicz, P., Eriksen, H.K., Banday, A.J., G\'orski, K.M., \& Lilje, P.B.
   2005, \apj, 635, 750
\bibitem[Bielewicz et al.(2004)]{bie04}
   Bielewicz, P., G\'orski, K.M., \& Banday, A.J. 2004, \mnras, 355, 1283
\bibitem[Brandt et al.(1994)]{bra94}
   Brandt, W.N., Lawrence, C.R., Readhead, A.C.S., Pakianathan, J.N.,
   \& Fiola, T.M. 1994, \apj, 424, 1
\bibitem[Chiang et al.(2003)]{chi03}
   Chiang, L.-Y., Naselsky, P.D., Verkhodanov, O.V., \& Way, M.J. 2003,
   \apj, 590, 65
\bibitem[Coles et al.(2004)]{col04}
   Coles, P., Dineen, P., Earl, J., \& Wright, D. 2004, \mnras, 350, 989
\bibitem[Copi et al.(2004)]{cop04}
   Copi, C.J., Huterer, D., \&  Starkman, G.D. 2004, \prd, 70, 043515
\bibitem[Copi et al.(2006)]{cop06}
   Copi, C.J., Huterer, D., Schwarz, D.J., \&  Starkman, G.D. 2006, \mnras, 
   367, 79
\bibitem[Cruz et al.(2005)]{cru05}
   Cruz, M., Mart\'{\i}nez-Gonz\'alez, E., Vielva, P., \& Cay\'on, L. 2005,
   \mnras, 356, 29
\bibitem[Cruz et al.(2006)]{cru06}
   Cruz, M., Tucci, M., Mart\'{\i}nez-Gonz\'alez, E., \& Vielva, P. 2006,
   \mnras, 369, 57
\bibitem[de Oliveira-Costa et al.(2004)]{deo04}
   de Oliveira-Costa, A., Tegmark, M., Zaldarriaga, M., \& Hamilton, A. 2004,
   \prd, 69, 063516
\bibitem[de Oliveira-Costa \& Tegmark(2006)]{deo06}
   de Oliveira-Costa, A., \& Tegmark, M. 2006, \prd, 74, 023005
\bibitem[Efstathiou(2004)]{efs04}
   Efstathiou, G. 2004, \mnras, 348, 885
\bibitem[Eriksen et al.(2004a)]{eri04:ilc}
   Eriksen, H.K., Banday, A.J., G\'orski, K.M., \& Lilje, P.B. 2004a,
   \apj, 612, 633
\bibitem[Eriksen et al.(2004b)]{eri04:asym}
   Eriksen, H.K., Hansen, F.K., Banday, A.J., G\'orski, K.M., \& Lilje, P.B.
   2004b, \apj, 605, 14
\bibitem[Eriksen et al.(2004c)]{eri04:genus}
   Eriksen, H.K., Novikov, D.I., Lilje, P.B., Banday, A.J., \& G\'orski, K.M.
   2004c, \apj, 612, 64
\bibitem[Freeman et al.(2006)]{fre06}
   Freeman, P.E., Genovese, C.R., Miller, C.J., Nichol, R.C., \& Wasserman, L.
   2006, \apj, 638, 1
\bibitem[G\'orski et al.(1999)]{gor99}
   G\'orski, K.M., Hivon, E., \& Wandelt, B.D. 1999,
   in Proceedings of the MPA/ESO Conference on "Evolution of Large-Scale 
   Structure: From Recombination to Garching", 
   ed. A.J. Banday, R.K. Sheth, \& L.N. da Costa, (Printpartners Ipskamp, NL),
   pp. 37-42 
\bibitem[G\'orski et al.(2005)]{gor05}
   G\'orski, K.M., Hivon, E., Banday, A.J., Wandelt, B.D., Hansen, F.K.,
   Reinecke, M., \& Bartelmann, M. 2005, \apj, 622, 759
\bibitem[Hinshaw et al.(2003)]{hin03}
   Hinshaw, G., et al. 2003, \apjs, 148, 135
\bibitem[Hinshaw et al.(2006)]{hin06}
   Hinshaw, G., et al. 2006, \apj, submitted (astro-ph/0603451)
\bibitem[Kogut et al.(2003)]{kog03}
   Kogut, A., et al. 2003, \apjs, 148, 161
\bibitem[Komatsu et al.(2003)]{kom03}
   Komatsu, E., et al. 2003, \apjs, 148, 119
\bibitem[Land \& Magueijo(2005)]{lan05}
   Land, K., \& Magueijo, J. 2005, \prl, 95, 071301
\bibitem[Naselsky et al.(2006)]{nas06}
   Naselsky, P.D., Novikov, I.D., \& Chiang, L.-Y. 2006, \apj, 642, 617
\bibitem[Park(2004)]{par04}
   Park, C.-G. 2004, \mnras, 349, 313
\bibitem[Page et al.(2003)]{pag03}
   Page, L., et al. 2003, \apjs, 148, 233
\bibitem[Page et al.(2006)]{pag06}
   Page, L., et al. 2006, \apj, submitted (astro-ph/0603450)
\bibitem[Peiris et al.(2003)]{pei03}
   Peiris, H.V., et al. 2003, \apjs, 148, 213
\bibitem[Schwarz et al.(2004)]{sch04}
   Schwarz, D.J, Starkman, G.D., Huterer, D., \& Copi, C.J. \prl, 93, 221301
\bibitem[Slosar \& Seljak(2004)]{slo04a}
   Slosar, A., \& Seljak, U. 2004, \prd, 70, 083002
\bibitem[Slosar, Seljak, \& Makarov(2004)]{slo04b}
   Slosar, A., Seljak, U., \& Makarov, A. 2004, \prd, 69, 123003
\bibitem[Spergel et al.(2003)]{spe03}
   Spergel, D.N., et al. 2003, \apjs, 148, 175
\bibitem[Spergel et al.(2006)]{spe06}
   Spergel, D.N., et al. 2006, \apj, submitted (astro-ph/0603449)
\bibitem[Tegmark et al.(2003)]{teg03}
   Tegmark, M., de Oliveira-Costa, A., \& Hamilton, A.J.S. 2003, \prd,
   68, 123523
\bibitem[Tegmark \& Efstathiou(1996)]{teg96}
   Tegmark, M., \& Efstathiou, G. 1996, \mnras, 281, 1297
\bibitem[Tojeiro et al.(2006)]{toj06}
   Tojeiro, R., Castro, P.G., Heavens, A.F., \& Gupta, S. 2006, \mnras,
   365, 265
\bibitem[Vielva et al.(2004)]{vie04}
   Vielva, P., Mart\'{\i}nez-Gonz\'alez, E., Barreiro, R.B., Sanz, J.L.,
   \& Cay\'on, L. 2004, \apj, 609, 22
\end{thebibliography}
\end{document}